\RequirePackage{fix-cm}

\documentclass[smallextended]{svjour3}       

\smartqed  

\makeatletter
\def\cl@chapter{\@elt {theorem}}
\makeatother

\usepackage{verbatim}

\usepackage{amsmath,amssymb,amsfonts}
\usepackage{algorithmic}
\usepackage{textcomp}

\usepackage{graphicx}
\usepackage{caption}
\usepackage{subcaption}
\usepackage{xcolor}
\usepackage{enumitem}
\usepackage{flushend}\setdescription{style=sameline,font=\normalfont\slshape}
\usepackage{tcolorbox}
\usepackage{soul}
\usepackage{color}
\usepackage{tikz}

\usepackage{cite}
\usepackage{hyperref}
\hypersetup{pdfborder=0 0 0}
\usepackage{cleveref}

\usepackage{tabularx}
\usepackage{booktabs}
\usepackage{multirow}
\usepackage{rotating}
\newenvironment{highlightbox}[1]{
    \begin{tcolorbox}[title={#1}]
    }{
    \end{tcolorbox}
}

\definecolor{hlreva}{RGB}{163, 255, 247}
\definecolor{hlrevb}{RGB}{178, 255, 163}
\definecolor{hlrevc}{RGB}{255, 163, 243}
\definecolor{hlrevd}{RGB}{252, 221, 134}

\journalname{Empirical Software Engineering}

\begin{document}

\title{Applying Bayesian Data Analysis for Causal Inference about Requirements Quality: A {Controlled} Experiment}
\titlerunning{Bayesian Data Analysis on Requirements Quality}

\author{Julian Frattini \and Davide Fucci \and Richard Torkar \and Lloyd Montgomery \and Michael Unterkalmsteiner \and Jannik Fischbach \and Daniel Mendez}
\authorrunning{Frattini et al.}

\institute{J. Frattini, D. Fucci, M. Unterkalmsteiner, and D. Mendez \at
                Blekinge Institute of Technology, Valhallavägen 1, 37140 Karlskrona, Sweden \\
                \email{\{firstname\}.\{lastname\}@bth.se}
            \and
                R. Torkar \at
                Chalmers and University of Gothenburg, 41756 Göteborg, Sweden \\
                Stellenbosch Institute for Advanced Study (STIAS), Stellenbosch, South Africa\\
                \email{richard.torkar@gu.se}
            \and
                L. Montgomery \at
                University of Hamburg, Mittelweg 177, 20148 Hamburg, Germany \\
                \email{lloyd.montgomery@uni-hamburg.de}
            \and J. Fischbach \at
                Netlight Consulting GmbH, Prannerstraße 4, 80333 München, Germany \\
                \email{jannik.fischbach@netlight.com}
            \and 
                J. Fischbach, D. Mendez \at 
                fortiss GmbH, Guerickestraße 25, 80805 München, Germany \\
                \email{\{lastname\}@fortiss.org}
}

\date{Received: date / Accepted: date}

\maketitle

\begin{tikzpicture}[remember picture,overlay]
    \node[anchor=south,yshift=10pt] at (current page.south) {\fbox{\parbox{1.5\textwidth-\fboxsep-\fboxrule\relax}{
        This version of the article has been accepted for publication, after peer review (when applicable) but is not the Version of Record and does not reflect post-acceptance improvements, or any corrections. The Version of Record is available online at: \url{http://dx.doi.org/10.1007/s10664-024-10582-1}
    }}};
\end{tikzpicture}

\begin{abstract} 
    It is commonly accepted that the quality of requirements specifications impacts subsequent software engineering activities.
    However, we still lack empirical evidence to support organizations in deciding whether their requirements are good enough or impede subsequent activities.
    We aim to contribute empirical evidence to the effect that requirements quality defects have on a software engineering activity that depends on this requirement.
    We {conduct} a controlled experiment in which 25 participants from industry and university generate domain models from four natural language requirements containing different quality defects.
    We evaluate the resulting models using both frequentist and Bayesian data analysis.
    Contrary to our expectations, our results show that the use of passive voice only has a minor impact on the resulting domain models.
    The use of ambiguous pronouns, however, shows a strong effect on various properties of the resulting domain models.
    Most notably, ambiguous pronouns lead to incorrect associations in domain models.
    Despite being equally advised against by literature and frequentist methods, the Bayesian data analysis shows that the two investigated quality defects have vastly different impacts on software engineering activities and, hence, deserve different levels of attention.
    Our employed method can be further utilized by researchers to improve reliable, detailed empirical evidence on requirements quality.
    \keywords{Requirements Engineering \and Requirements Quality \and Experiment \and Replication \and Bayesian Data Analysis}
\end{abstract}

\section{Introduction}
\label{sec:intro}

Software requirements specify the needs and constraints that stakeholders impose on a desired system. 
Software requirements specifications (SRS), the explicit manifestation of requirements as an artifact~\cite{mendez2019artefacts}, serve as input for various subsequent software engineering (SE) activities, such as deriving a software architecture, implementing features, or generating test cases~\cite{mendez2015artefact}.
As a consequence, the quality of an SRS impacts the quality of \textit{requirements-dependent} activities~\cite{frattini2023requirements,femmer2018requirements,femmer2015ABREQM}. 
A quality defect in an SRS---for example, an ambiguous formulation---can cause differing interpretations and result in the design and implementation of a solution that does not meet the stakeholders' needs~\cite{mendez2017naming}.
The inherent complexity of natural language (NL), which is most commonly used for specifying requirements~\cite{franch2017practitioners}, aggravates this challenge further.
Since quality defects are understood to scale in cost for removal~\cite{boehm1984software}, organizations are interested in identifying and removing these defects as early as possible~\cite{montgomery2022empirical}.

Within the requirements engineering (RE) research domain, the field of requirements quality research aims to meet this challenge~\cite{montgomery2022empirical}.
Requirements quality research has already identified several attributes of requirements quality~\cite{montgomery2022empirical} (e.g., unambiguity, completeness, consistency) and proposes \textit{quality factors}, i.e., requirements writing rules (e.g., the use of \textit{passive voice} being associated with bad quality~\cite{femmer2014impact}) as well as tools that automatically detect alleged quality defects~\cite{femmer2017rapid}.
However, existing approaches fall short in at least three regards~\cite{frattini2023requirements}: 
i) only a fraction of publications provide empirical evidence that would demonstrate the impact of quality defects~\cite{montgomery2022empirical}, 
ii) the few empirical studies that do so largely ignore potentially confounding context factors~\cite{mund2015does,juergens2010much}, and 
iii) the analyses conducted in existing publications do not go beyond binary insights (i.e., a quality factor \textit{does} have an impact or it \textit{does not})~\cite{femmer2014impact,deissenboeck2007activity}. 
These gaps have impeded the adoption of requirements quality research in practice~\cite{franch2017practitioners}.

In this article, we aim to address the above-mentioned shortcomings by
i) {conducting a controlled experiment with 25 participants simulating a requirements-dependent activity (i.e., domain modeling) using four natural-language requirements as input.
The experiment contributes empirical evidence on the impact of two commonly researched quality factors \textit{passive voice}~\mbox{\cite{femmer2014impact}} and \textit{ambiguous pronouns}~\mbox{\cite{ezzini2022automated}}.
The investigation of the impact of passive voice is a conceptual replication~\mbox{\cite{baldassarre2014replication}} of the only controlled experiment studying the impact of passive voice on domain modeling~\mbox{\cite{femmer2014impact}} known to us.
Therefore, our experiment also strengthens the robustness of their conclusions by providing diagnostic evidence~\mbox{\cite{nosek2020replication}}. 
}
Further, we ii) collect data about relevant context factors such as experience in software engineering (SE) and RE, domain knowledge, and task experience, and integrate these data in our data analysis.
Finally, we iii) contrast the state-of-the-art frequentist data analysis (FDA) with Bayesian data analysis (BDA), {which entails both a causal framework and Bayesian modeling for statistical causal inference~\mbox{\cite{mcelreath2020statistical}}.}
The latter has recently been popularized in SE research~\cite{furia2019bayesian} since it generates more nuanced empirical insights.
Our study is categorized as a laboratory experiment in a contrived setting~\cite{stol2018abc}, isolating the effect of the selected quality factors of interest. 
The causal inference of their impact contributes to our long-term goal of providing an empirically grounded understanding of the impact of requirements quality.
This will support organizations in assessing their requirements and detecting relevant quality defects early.


This paper makes the following contributions:

\begin{enumerate}
   \item {a controlled experiment investigating the impact of requirements quality;} 
   \item {a conceptual replication of the only controlled experiment investigating the impact of passive voice~\mbox{\cite{femmer2014impact}};}
   \item the application of BDA to requirements quality research, which is among the first of its kind in RE; and
   \item {an archived replication package containing all supplementary material, including protocols and guidelines for data collection and extraction, the raw data, analysis scripts, figures, and results~\mbox{\cite{frattini2024replication}}.}
\end{enumerate}

The remainder of this manuscript is organized as follows. 
\Cref{sec:related} introduces relevant related work. 
We present our research method in \Cref{sec:method} and the results in \Cref{sec:results}.
We discuss these results in \Cref{sec:discussion} before concluding our manuscript in \Cref{sec:conclusion}.

\section{Background}
\label{sec:related}

\Cref{sec:related:quality} introduces the research domain of this work by summarizing existing research on requirements quality.
\Cref{sec:related:bdainse} motivates BDA---the statistical tool employed in this work---by explaining its adoption in SE research.

\subsection{Requirements Quality}
\label{sec:related:quality}

\Cref{sec:related:quality:general} introduces the general area of requirements quality research and \Cref{sec:related:quality:examples} presents two research directions within.
\Cref{sec:related:quality:shortcomings} summarizes the three major shortcomings that currently challenge requirements quality research.

\subsubsection{Requirements Quality Research}
\label{sec:related:quality:general}

It is commonly accepted that the quality of requirements specifications impacts subsequent SE activities, which depend on these specifications~\cite{femmer2018requirements,frattini2023requirements}.
Quality defects in requirements specifications may, therefore, ultimately cause budget overrun~\cite{philippo2013requirement} or even project failure~\cite{mendez2017naming}.
Two further factors aggravate the effect.
Firstly, natural language (NL), which is inherently ambiguous and, hence, prone to quality defects, remains the most commonly used syntax to specify requirements~\cite{franch2023state,wagner2019status}.
Secondly, the cost of removing quality defects scales the longer they remain undetected~\cite{boehm1984software}.
For example, clarifying an ambiguous requirements specification takes comparatively less effort than re-implementing a faulty implementation based on the ambiguous specification.
However, it requires detecting the ambiguity and predicting that the ambiguity potentially causes the implementation to become faulty before it happens.
These circumstances necessitate managing the quality of requirements specifications to detect and remove requirements quality defects preemptively.

Requirements quality research seeks answers to this need~\cite{montgomery2022empirical}.
One main driver of this research is \textit{requirements quality factors}~\cite{frattini2022live}, i.e., metrics that can be evaluated on NL requirements specifications to determine quality defects.
For example, the \textit{voice} of an NL sentence (active or passive) is considered a quality factor, as the use of \textit{passive voice} is associated with bad requirements quality due to potential omission of information~\cite{femmer2014impact}.
Automatic detection techniques using natural language processing (NLP)~\cite{zhao2021natural} can automatically evaluate quality factors to detect defects in NL requirements specifications~\cite{femmer2017rapid}.

\subsubsection{Existing Research on Passive Voice and Ambiguous Pronouns}
\label{sec:related:quality:examples}

We present two examples of commonly researched requirements quality factors in the following sections.

\paragraph{Passive Voice}

\begin{figure}
    \centering
    \includegraphics[width=0.8\textwidth]{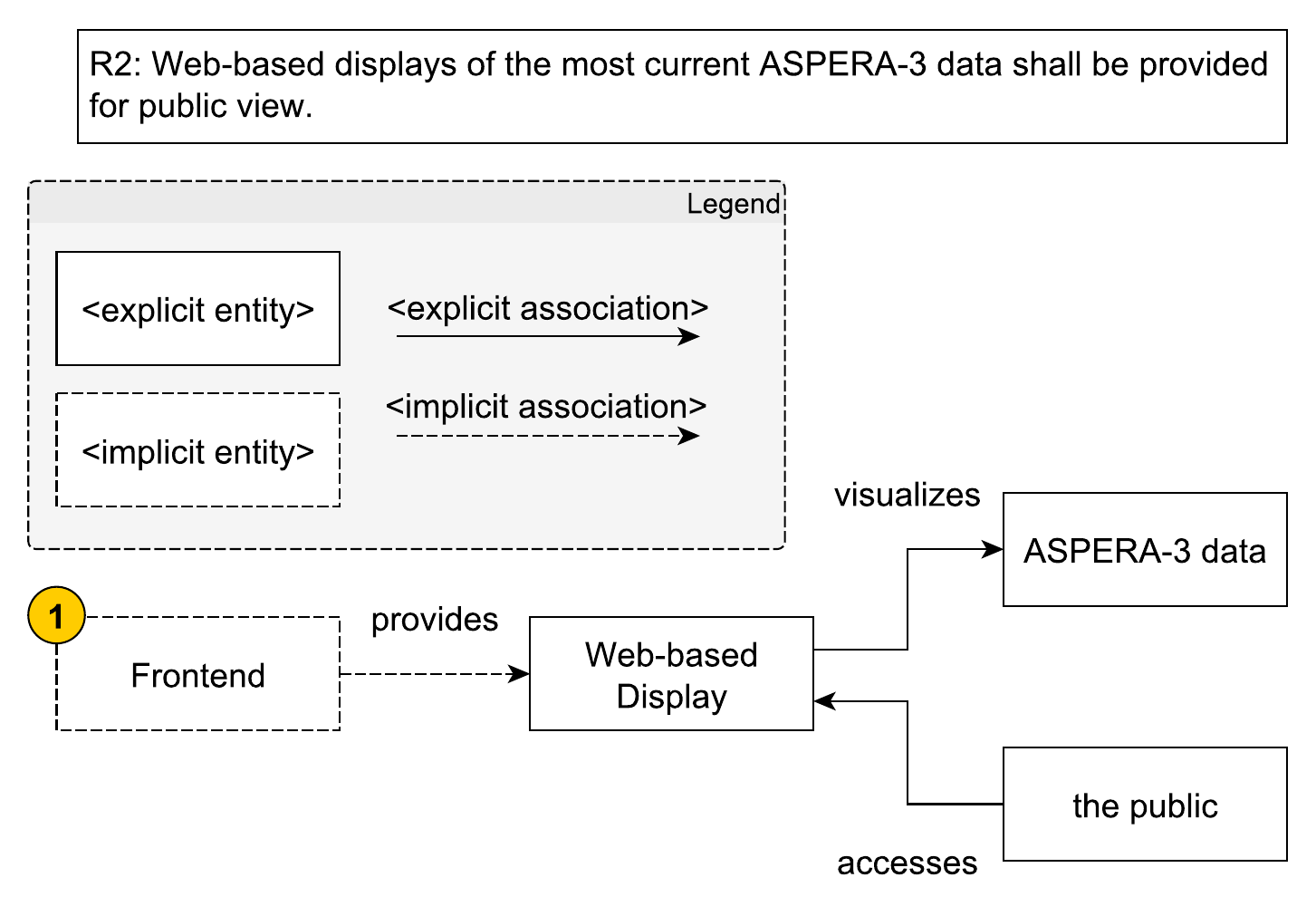}
    \caption{Formalization of a requirements specification R2 using passive voice}
    \label{fig:example:passivevoice}
\end{figure}

One commonly researched requirements quality factor is using \textit{passive voice} in natural language requirements specifications. 
A sentence in passive voice elevates the semantic patient rather than the semantic agent of the main verb to the grammatical subject~\cite{ogrady2001linguistics}.
For example, in the passive voice sentence ``Web-based displays of the most current ASPERA-3 data shall \textbf{be provided} for public view.'', the patient of the \textit{providing} process---the ``web-based displays''---becomes the grammatical subject of the sentence.
Even though passive voice sentences may still contain the semantic agent (e.g., ``Web-based displays of the most current ASPERA-3 data shall be provided for public view \textit{by a front-end}.''), writers often omit it intentionally or unintentionally~\cite{krisch2015myth}.
\Cref{fig:example:passivevoice} visualizes the omission of the semantic agent in this exemplary requirement specification.

Omitting the semantic agent of a sentence in a passive voice formulation obscures critical information in a requirements specification.
Hence, requirements quality guidelines advise against using passive voice~\cite{pohl2016requirements}.
However, while several guidelines advise against the use of passive voice based on the theoretical argument of information omission presented above~\cite{pohl2016requirements,genova2013framework,gleich2010ambiguity,kof2007treatment,hasso2019detection}, only two papers investigate whether passive voice has an actual impact on requirements quality:
Krisch et al.\ let domain experts rate active and passive voice requirements as either problematic or unproblematic.
They concluded that most passive voice requirements were unproblematic as the surrounding context information compensated the omission of the semantic agent of the sentence~\cite{krisch2015myth}.
Femmer et al.\ conducted an empirical investigation of the impact of the use of passive voice in requirements specification on the domain modeling activity in a controlled experiment.
They concluded that passive voice only causes missing relationships from the domain model, but not missing actors or entities as initially assumed~\cite{femmer2014impact}.
The limited evidence for the harmfulness of using passive voice in requirements specifications~\cite{krisch2015myth,femmer2014impact} stands in stark contrast to the amount of tools and approaches proposed to automatically detect quality defects by identifying the use of passive voice~\cite{femmer2017rapid,kof2007treatment,knauss2009learning,femmer2018requirements,hasso2019detection,rosadini2017using,soeken2014quality,drechsler2014automated,parra2015methodology,ferrari2018detecting,genova2013framework}.

\paragraph{Ambiguous Pronouns}

\begin{figure}
    \centering
    \includegraphics[width=0.8\textwidth]{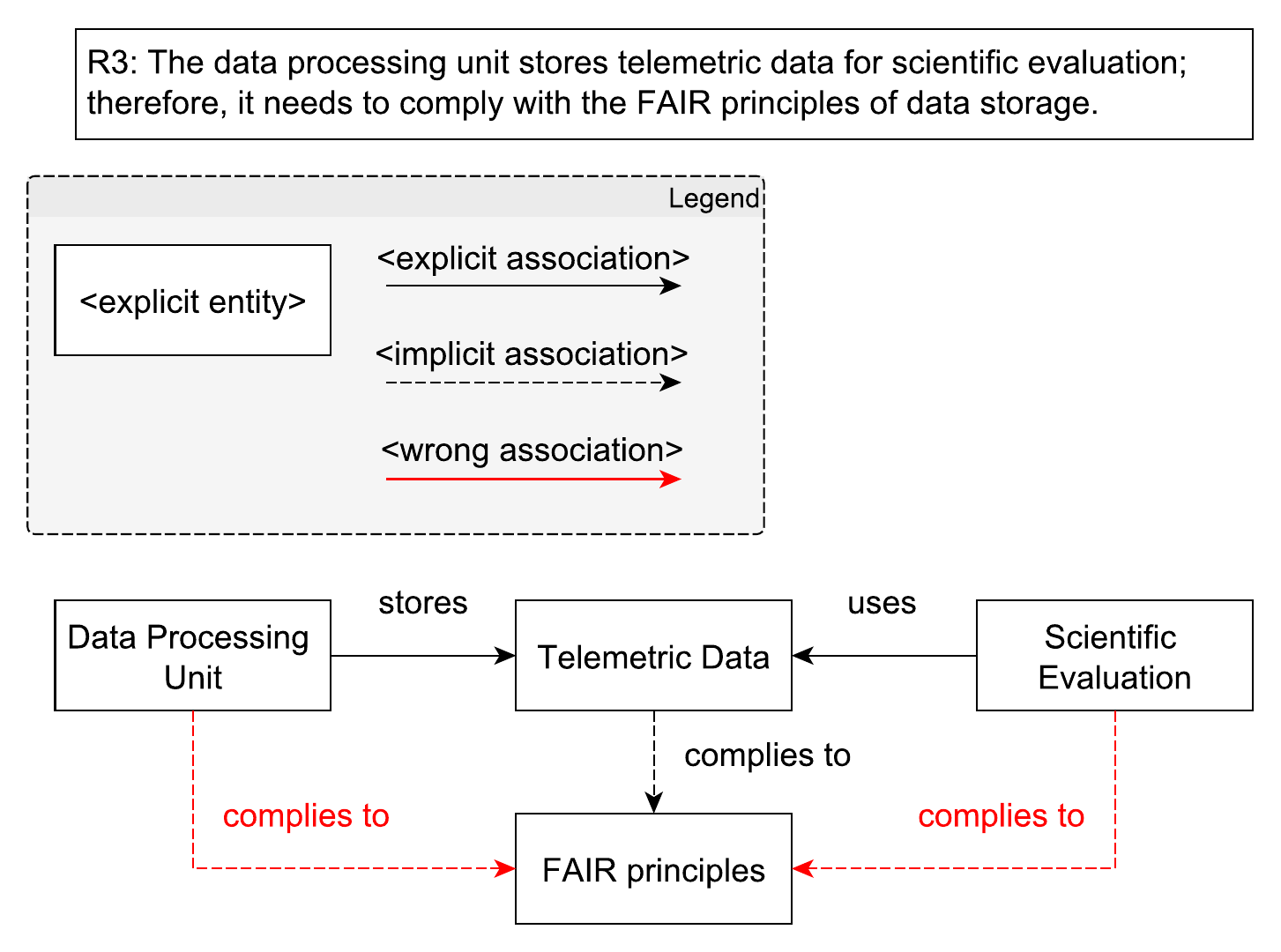}
    \caption{Formalization of a requirements specification R3 using an ambiguous pronoun}
    \label{fig:example:ambiguouspronoun}
\end{figure}

The inherent ambiguity of natural language~\cite{poesio1996ambiguity} poses several challenges for requirements specifications using natural language~\cite{nuseibeh2000requirements,bano2015addressing}.
One commonly researched requirements quality factor related to ambiguity is the use of \textit{ambiguous pronouns}, which is a type of \textit{referential ambiguity}~\cite{berry2004ambiguity}.
An ambiguous pronoun exhibits \textit{anaphoric ambiguity}, that ``occurs when a pronoun can plausibly refer to different entities and thus be interpreted differently by different readers''~\cite{ezzini2022automated}.
For example, in the requirements specification ``The \textit{data processing unit} stores \textit{telemetric data} for \textit{scientific evaluation}; therefore, \textbf{it} needs to comply with the FAIR principles of data storage.'', the pronoun \textit{it} could syntactically refer to the ``data processing unit'', the ``telemetric data'', or the ``scientific evaluation.''
\Cref{fig:example:ambiguouspronoun} visualizes how a reader can resolve the reference.

To avoid deviating interpretations of a requirements specification, established requirements quality guidelines advise against the use of ambiguous pronouns~\cite{pohl2016requirements} at the expense of conciseness.
However, the number of publications proposing tools and algorithms to automatically identify and resolve ambiguous pronouns~\cite{ezzini2022automated,yang2010extending,yang2011analysing,kamsties2000taming,sharma2016machine,ezzini2022taphsir,shah2015resolving,ezzini2021using,chantree2006identifying} significantly outweighs the singular publication that actually has empirically investigated the effect of ambiguous pronouns.
Kamsties et al.\ investigated the effects of formalizing requirements, which included evaluating the propagation of ambiguous pronouns from NL into more formal specifications~\cite{kamsties2005empirical}.
Their experiment involving students revealed that 20-37\% of all ambiguous pronouns were incorrectly resolved while formalizing NL requirements specifications.
While Kamsties et al.\ concluded that requirements formalization does not sufficiently resolve ambiguities, these results also support the assumption that ambiguous pronouns propagate into subsequent artifacts depending on the requirements specifications.
On the contrary, the scarce empirical work on the effect of ambiguity in general (not specifically ambiguous pronouns) agrees that ambiguity has a negligible effect on downstream software engineering activities~\cite{de2010ambiguity,philippo2013requirement}.
Other than these empirical contributions, the aforementioned publications proposing solutions rather than investigating the relevance of the problem refer to deontic guidelines~\cite{pohl2016requirements,belev1989guidelines}, anecdotal evidence about ambiguity in general~\cite{boyd2005measuring,firesmith2007common,ferrari2019nlp,christel1992issues}, or---in very rare cases---cognitive science theory~\cite{poesio1996ambiguity}.

\subsubsection{Shortcomings in Requirements Quality Research}
\label{sec:related:quality:shortcomings}

The previous examples highlight at least three shortcomings from which requirements quality research suffers.

\paragraph{Lack of empirical evidence}

First, the relevance of quality factors like passive voice or ambiguous pronouns is rarely determined empirically~\cite{frattini2023requirements}.
Scientific contributions proposing solutions (i.e., detecting or removing quality defects) outweigh those investigating the actual extent of the assumed problem.
Without knowledge about this extent, it remains unclear whether a proposed solution addresses a problem that is actually relevant to practice.

Previous systematic research has come to the same conclusion.
For example, in a previous systematic study, we determined that the effect of quality defects is determined empirically in only 18\% of the publications included in our sample~\cite{frattini2023requirements}.
Bano et al.\ found only two publications within their sample of 28 studies that empirically investigated the importance of ambiguity detection~\cite{bano2015addressing}.
Montgomery et al.\ systematically investigated empirical research on requirements quality research and also concluded that most studies focus on improving requirements quality (i.e., detecting and removing defects) rather than defining or evaluating it (i.e., understanding the actual effect)~\cite{montgomery2022empirical}.
Instead, most requirements quality publications draw on anecdotal evidence and unproven hypotheses~\cite{frattini2023requirements}.
This lack of empirical evidence undermines the trust in requirements quality research and hinders its adoption in practice~\cite{franch2017practitioners,phalp2007assessing,femmer2018scout}. 

\paragraph{Lack of context}

Second, existing research mostly ignores the influence of context factors on the effect of quality defects~\cite{frattini2023requirements}.
Context factors encompass all human and organizational factors influencing the downstream SE activities involving requirements specification~\cite{petersen2009context}.
For example, the \textit{domain experience} of a stakeholder or the \textit{process model} used during development may mediate the effect of ambiguity in requirements specifications~\cite{philippo2013requirement}.

Requirements quality research has acknowledged the relevance of context factors to requirements quality~\cite{mund2015does,juergens2010much}.
Recent propositions have advocated for a shift away from the unrealistic goal of developing a one-size-fits-all solution to requirements quality and, instead, moving towards more context-sensitive research~\cite{briand2017case,mendez2017naming}.
However, this initiative has shown little effect in requirements quality research so far~\cite{frattini2023requirements}.

\paragraph{Lack of detailed projections}

Third, the few empirical contributions to requirements quality research limit their insights to categorical projections, i.e., the evaluation of a quality factor on a requirements specification (e.g., \textit{using passive voice} or \textit{not using passive voice}) are projected on a categorical scale (e.g., \textit{good quality} or \textit{bad quality}).
Most commonly, the categorical output space consists of two~\cite{deissenboeck2007activity} (\textit{impact} or \textit{no impact}) or three~\cite{femmer2015ABREQM} (\textit{positive impact}, \textit{no impact}, or \textit{negative impact}) categories.
This simplification inhibits a nuanced comparison of different quality factors.
On an absolute scale, a quality factor having an impact does not automatically entail that this impact is significant and warrants resources for detection and mitigation.
On a relative scale, two quality factors that have an impact are impossible to compare to allocate resources towards the more significant one.
Consequently, even empirical contributions to the field of requirements quality lack sophisticated insights that would support organizations in determining and dealing with relevant quality factors to control during the RE phase.

\begin{highlightbox}{Requirements Quality Research Gaps}
    Requirements quality research suffers from (1) a lack of empirical evidence about the relevance of quality factors, (2) a lack of context-sensitivity, and (3) evaluations of impact that are more fine-grained than categorical.
\end{highlightbox}

\subsubsection{Requirements Quality Theory}
\label{sec:related:quality:rqt}

\begin{figure}
    \centering
    \includegraphics[width=\textwidth]{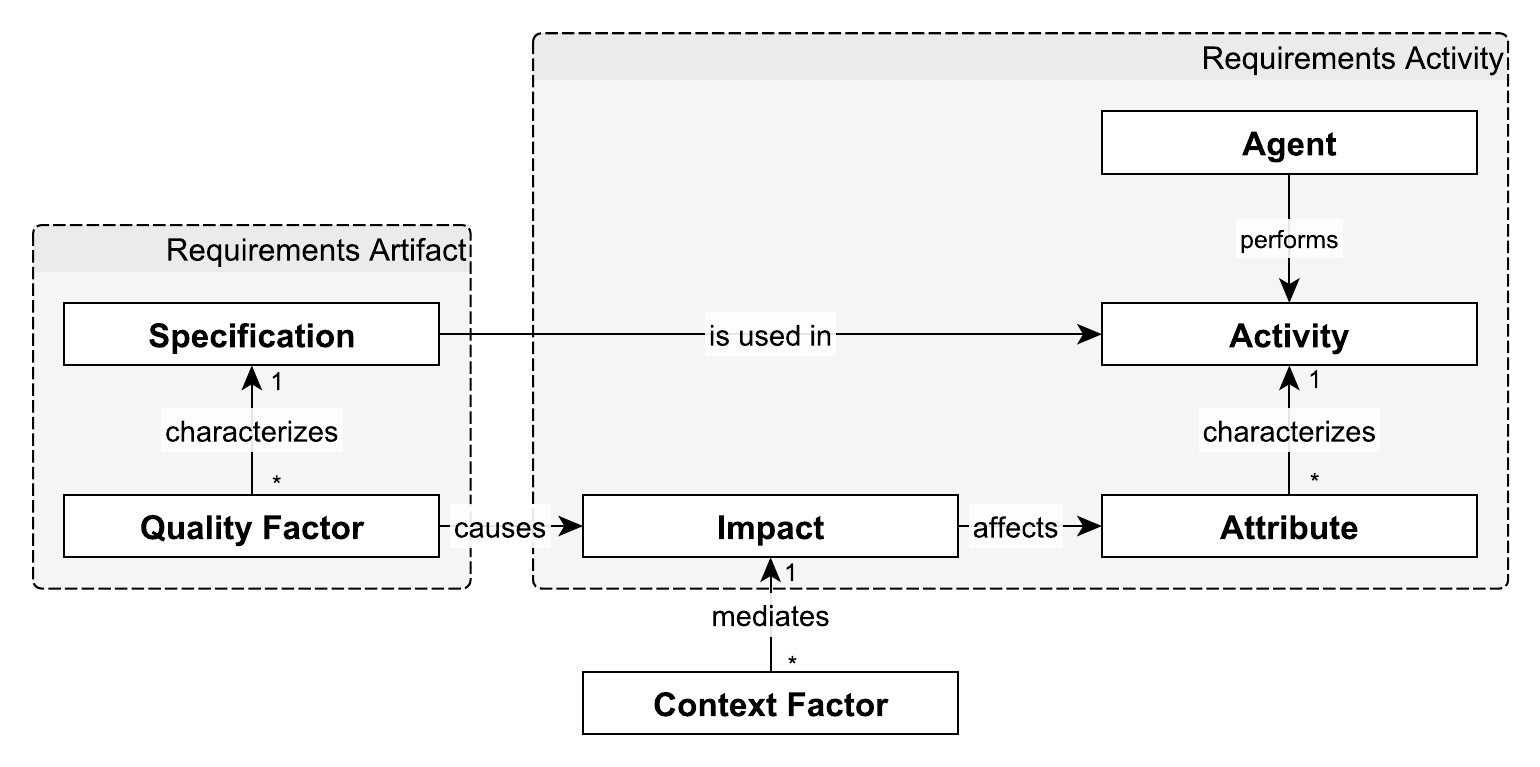}
    \caption{Reduced version of the activity-based Requirements Quality Theory~\cite{frattini2023requirements}}
    \label{fig:rqt}
\end{figure}

Based on the identification of the above-mentioned shortcomings, we have developed a requirements quality theory in previous research~\cite{frattini2023requirements}.
This theory frames requirements quality as the impact that properties of requirements specifications (called the \textit{quality factors}) in combination with context factors have on the properties (called \textit{attributes}) of activities that use these specifications as input.
\Cref{fig:rqt} visualizes the requirements quality theory.

The requirements quality theory facilitates overcoming the aforementioned shortcomings.
Because the RQT makes the quality of a requirements specification dependent on its impact on subsequent activities, it demands empirical evidence about this impact before claiming that a quality factor reflects actual requirements quality.
The inclusion of context factors in the definition of requirements quality mandates context-sensitivity.
The abstraction of the impact concept allows for more advanced relationships between specifications and impacted activities than just the categorical type.

However, while the requirements quality research draws on mature software quality research~\cite{deissenboeck2007activity,wagner2012quamoco}, it has not been actively used yet.
Even the predecessor of the theory~\cite{femmer2015ABREQM} was explicitly ignored in follow-up research by its authors due to the complexity of its implementation~\cite{femmer2017rapid}.
The work presented in this manuscript constitutes the first application of the theory known to the authors.

\subsection{Bayesian data analysis in software engineering}
\label{sec:related:bdainse}

In recent years, SE research has adopted Bayesian data analysis (BDA) for {statistical causal inference}.
BDA signifies a departure from frequentist methods like null-hypothesis significance testing (NHST), the previous state-of-the-art in terms of inferential statistics in SE research.
NHST determines whether there is a ``statistically significant'' difference between two or more distributions.
Observations of a dependent variable are stratified by an independent variable to obtain a binary answer of whether or not different values of the independent variable correlate with different distributions of the dependent variable.

{Opposed to that, BDA encourages the use of causal frameworks~\mbox{\cite{mcelreath2020statistical}}.
These frameworks make causal assumptions explicit~\mbox{\cite{elwert2013graphical}} and allow reasoning about causally relevant variables~\mbox{\cite{pearl1995bayesian,pearl2016causal}}.
Furthermore,} BDA abstains from reducing complex variable distributions to binary inference~\cite{mcelreath2020statistical}.
Instead, dependent variables are expressed as a probability distribution, which preserves the natural uncertainty with which any variable is determined.
Similarly, the impact of any independent variable on the dependent variable is expressed in terms of a probability distribution.
Using Bayes' Theorem, these assigned prior probability distributions are updated with observed data to obtain a posterior probability distribution~\cite{furia2019bayesian}.
Given the observed data, these posterior probability distributions model the most likely impact of variable values.
BDA methods are becoming widely adopted also due to the modern computational power enabling Markov Chain Monte Carlo (MCMC) randomized algorithms~\cite{brooks2011handbook}, tools like \texttt{Stan}~\cite{carpenter2017stan}, and libraries like \texttt{rethinking}~\cite{mcelreath2020statistical} and \texttt{brms}~\cite{burkner2017brms}.

While BDA is associated with a much steeper learning curve than frequentist methods, it offers several advantages.

\begin{enumerate}
    \item BDA is not based on the unsound probabilistic extension of the \textit{modus tollens} like frequentist hypothesis testing. The modus tollens ($P \rightarrow Q, \neg Q \therefore \neg P$, or \textit{if P implies Q and Q is false, then P must also be false}) applies to propositional, Boolean logic, but not when inferring from probabilities~\cite{furia2019bayesian}.
    \item BDA provides more complex insights than point-wise comparisons. Although BDA lacks out-of-the-box statistical methods like frequentists' t-tests that are simple to apply, its results reflect the uncertainty of the data, the influence of context, and they can be interpreted more intuitively.
    \item {The causal framework entailed by BDA} makes causal assumptions explicit. The Bayesian workflow~\cite{gelman2020bayesian} makes any hypothesis of causal relations explicit. Analyses become more transparent, and competing causal assumptions are easier to assess.
\end{enumerate}

Furia et al.~\cite{furia2019bayesian}, and Torkar et al.~\cite{torkar2020missing} advocate for the adoption of BDA in software engineering research by discussing its advantages over the frequentist counterpart and mitigating its steep learning curve with extensive demonstrations~\cite{furia2022applying}.
SE researchers have begun to apply BDA in various evaluations.
Previous studies have used BDA to model bug-fixing time in open source software projects~\cite{vieira2022bayesian}, to confirm the broken window theory in SE~\cite{leven2022broken}, to investigate gender differences in personality traits of software engineers~\cite{russo2020gender}, and to understand data-driven decision making practices~\cite{svensson2019unfulfilled}.
In the area of requirements engineering, BDA has been used to evaluate the effect of obsolete requirements on software estimation~\cite{gren2021possible} and to compare requirements prioritization criteria~\cite{berntsson2021not}.

\section{Method}
\label{sec:method}

We {conducted} a controlled experiment that investigates the impact of requirements quality on a software engineering activity.
Our goal is both to (1) contribute empirical evidence to the effect of quality defects and (2) compare the inferential capabilities of frequentist (FDA) with Bayesian (BDA) statistics.
{Part of our experiment contributes a conceptual replication~\mbox{\cite{baldassarre2014replication}} of the study conducted and reported by Femmer et al.~\mbox{\cite{femmer2014impact}} {and re-analyzed by us~\mbox{\cite{frattini2024second}}}, as a subset of our hypotheses overlaps with theirs and our study contributes diagnostic evidence for their claims~\mbox{\cite{nosek2020replication}}. 
Therefore, we report the design of the experiment with emphasis on the replication following the guidelines by Carver~\mbox{\cite{carver2010towards}}. 

\subsection{Goals}
\label{sec:method:goals}

We formulate our goal using the goal-question metric approach~\cite{wohlin2012experimentation}.
We aim to \textit{characterize} the impact of passive sentences and sentences using ambiguous pronouns in requirements on domain modeling \textit{with respect to} the quality of the created domain model artifacts \textit{from the point of view of} software engineers \textit{in the context of} an analysis of requirements from an industrial project.
In this definition, \textit{software engineer} includes all roles that work with requirements specifications, including software developers, requirements engineers, business analysts, managers, and more.
We derive the following research questions from our goal:

\begin{itemize}
    \item RQ1: Do quality defects in NL requirements specifications harm the domain modeling activity?
    \begin{itemize}
        \item RQ1.1: Does the use of \textit{passive voice} in NL requirements specifications harm the \textit{duration}, \textit{completeness}, \textit{conciseness}, and \textit{correctness} of the domain modeling activity?
        \item RQ1.2: Does the use of \textit{ambiguous pronouns} in NL requirements specifications harm the \textit{duration}, \textit{completeness}, \textit{conciseness}, and \textit{correctness} of the domain modeling activity? 
        \item RQ1.3: Does the combined use of \textit{passive voice} and \textit{ambiguous pronouns} in NL requirements specifications harm the \textit{duration}, \textit{completeness}, \textit{conciseness}, and \textit{correctness} of the domain modeling activity? 
    \end{itemize}
    \item RQ2: Do context factors influence the domain modeling activity?
    \begin{itemize}
        \item RQ2.1: Do context factors harm the domain modeling activity?
        \item RQ2.2: Do context factors mediate the impact of quality defects on the domain modeling activity?
    \end{itemize}
\end{itemize}

RQ1 is dedicated to the main relationship of interest between quality defects and an affected activity.
{RQ1.1 aligns with the research question driving the original study~\mbox{\cite{femmer2014impact}}, which makes this part of our study a conceptual replication.} 
RQ1.2 extends the scope of the investigation of quality factors with ambiguous pronouns.
RQ1.3 investigates the interaction between the two quality factors.
RQ2 adds a context-sensitive perspective to the relationship.
RQ2.1 focuses on the direct effect that context factors have on the affected activity.
RQ2.2 investigates whether context factors mediate the effect of quality defects on the activity.

\subsection{Original Experiment}
\label{sec:method:original}

The original study~\cite{femmer2014impact} addresses the research question ``Is the use of passive sentences in requirements harmful for domain modeling?''
The authors involved 15 university students from different study programs (2 B.Sc., 8 M.Sc., 4 Ph.D., one unknown) in a controlled randomized experiment with a parallel design~\cite{wohlin2012experimentation}.
Each participant was assigned to one of two groups and received seven requirements that were formulated either using active or passive voice. 
The experimental task was to derive a domain model from each requirement that contains all relevant actors, domain objects, and associations between them.
The study material and results are available online.\footnote{\url{https://doi.org/10.5281/zenodo.7499290}}

For the dependent variable, the authors calculated the \textit{number of missing domain model elements} (i.e., actors, objects, and associations). 
Although the authors also recorded context variables such as a categorical assessment of general knowledge in SE and RE, these were not used in the analysis.
The analysis followed a frequentist approach performing a null-hypothesis significance test for each of the three domain model elements to determine whether a statistically significant difference between the experimental groups exists. 
The study shows a statistically significant difference in the number of identified associations but not in the number of actors or objects.
The authors conclude that the commonly assumed impact of passive voice on missing domain model actors is actually negligible, but passive voice impedes the understanding of the relationships between entities in the requirements specification.

\subsection{{Reanalysis}}

{The original study by Femmer et al. analyzed its data under simplified assumptions. 
Among these is the assumption that the three dependent variables (number of missing actors, objects, and associations) only depend on the main factor (use of active or passive voice).
We challenged this assumption in a re-analysis of the original data~\mbox{\cite{frattini2024second}} for the following reasons:}

\begin{enumerate}
    \item {In a small-scale experiment employing a parallel design, there is no measure to control subject variability~\mbox{\cite{vegas2015crossover}}, such that context factors like experience or skill might affect the dependent variables.}
    \item {Missing an actor or object in the domain model (i.e., a node) necessarily causes an association to be missed (i.e., an edge that would have connected these nodes).}
\end{enumerate}

{\mbox{\Cref{fig:ipv:original}} visualizes the causal assumptions of the original experiment~\mbox{\cite{femmer2014impact}} as a directed acyclic graph~\mbox{\cite{elwert2013graphical}} (the syntax of which is further explained in \mbox{\Cref{sec:method:analysis:bda}}) and \mbox{\Cref{fig:ipv:revised}} shows the revision in scope of the reanalysis~\mbox{\cite{frattini2024second}}.
The revision includes (1) two context factors that were already recorded but not used in the original experiment, and (2) two causal relations between the response variables.}

\begin{figure}
    \centering
    \begin{subfigure}[b]{0.9\textwidth}
         \centering
         \includegraphics[width=\textwidth]{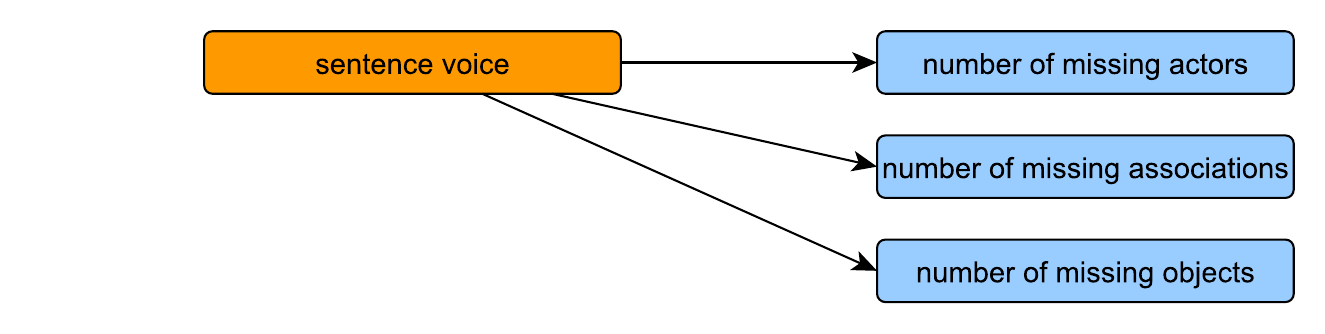}
         \caption{{Original causal assumptions~\mbox{\cite{femmer2014impact}}}}
         \label{fig:ipv:original}
     \end{subfigure}
     \begin{subfigure}[b]{0.9\textwidth}
         \centering
         \includegraphics[width=\textwidth]{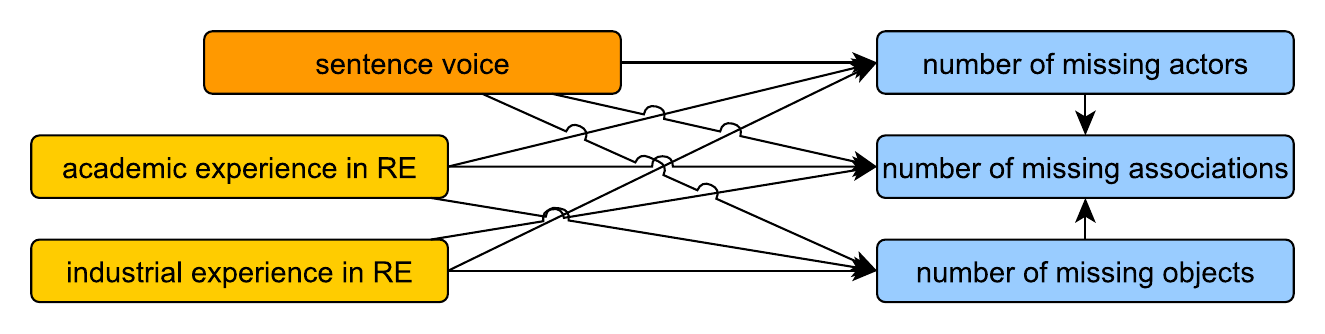}
         \caption{{Revised causal assumptions~\mbox{\cite{frattini2024second}}}}
         \label{fig:ipv:revised}
     \end{subfigure}
     \begin{subfigure}[b]{0.8\textwidth}
         \centering
         \includegraphics[width=\textwidth]{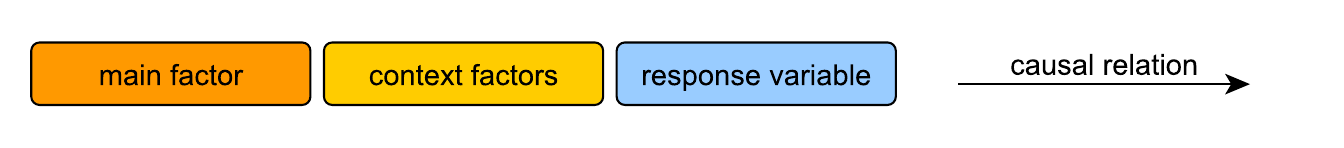}
     \end{subfigure}
     \caption{{Causal assumptions about the impact of passive voice}}
     \label{fig:ipv}
\end{figure}

{We performed a re-analysis, i.e., an independent analysis of the same data using a different statistical model~\mbox{\cite{gomez2010replications}}, which is sometimes referred to as a test of robustness~\mbox{\cite{nosek2020replication}}.
During this re-analysis, we replaced the NHSTs with regression models that include context factors and the affecting response variables in the case of missing associations.

The results of this re-analysis agree with the original study in that the effect of passive voice on the number of missing actors and objects is negligible.
However, the re-analysis disagrees with the original study regarding the effect of passive voice on the number of missing associations.
The re-analysis determined that passive voice slightly increases the number of missing associations ($\beta_{pv}=0.7$). 
Still, the confidence interval of this effect ($CI_{pv}=(-0.56, 1.90)$) intersects 0 and is, therefore, not significant.
On the other hand, the effect of missing objects on missing associations was significant ($\beta_{act}=1.12$, $CI_{act}=(0.32, 1.96)$).
The re-analysis did not find a significant effect of the available context variables on the response variables.
Our re-analysis concludes that the effect of passive voice on the domain modeling activity is less significant than originally assumed~\mbox{\cite{frattini2024second}}.
}

\subsection{Our Experiment}
\label{sec:method:own}

{The reanalysis~\mbox{\cite{frattini2024second}} of the study by Femmer et al.~\mbox{\cite{femmer2014impact}} did improve the conclusion validity of the results but failed to address other shortcomings.
For example, the subject variability still threatened the internal validity of the results due to the parallel design of the experiment~\mbox{\cite{vegas2015crossover}}, and the context factors were limited to those recorded during the original study.
Hence, we used their study as inspiration for our own presented in this paper and aimed to improve upon the research design.}
During the preparation of our study, we conferred with the authors of the original study and made the following changes to the original study.

\begin{itemize}
    \item \textbf{Experimental design}: We employ a factorial crossover instead of a parallel design, which minimizes the risk of confounding (i.e., each participant acts as their own control) while requiring a smaller sample~\cite{vegas2015crossover}.
    \item \textbf{Independent variables}: This study investigates---in addition to using passive voice---the impact of \textit{ambiguous pronouns} in requirements specifications and their combined usage to extend the range of requirements quality defects. 
    \item \textbf{Dependent variables}: We merged two types of elements in the domain model (the nodes of the model, i.e., \textit{actors} and \textit{objects}) into a single type \textit{entity} {because they represent the same concept in the domain model (nodes)~\mbox{\cite{femmer2014impact}} and the distribution in our experimental objects is heavily skewed (16/17 entities are objects).} Furthermore, we increased the dependent variables by additionally evaluating the number of \textit{superfluous entities}, the number of \textit{wrong associations}, and the \textit{duration} for creating the domain model.
    \item \textbf{Sampling strategy}: We sample from both students and practitioners of software engineering to more accurately represent the target population of software engineers. This change aims at increasing the external validity of our results~\cite{baltes2022sampling}.
    \item \textbf{Instrumentation}: The participants performed the experimental task online using a web-based application rather than offline using pen and paper. This allows for more flexibility in reaching industry participants~\cite{baltes2022sampling}.
    \item \textbf{Context factors}: To obtain a richer understanding of the impact of quality defects, we included seven additional context factors. This change made it necessary to extend the questionnaire used in the original study to collect demographic information from the participants.
    \item \textbf{Experimental object}: We sampled the objects from a data set of industrial requirements specifications~\cite{ferrari2017pure} rather than from a requirements specification written in a student project~\cite{femmer2014impact} to increase the realism of the experimental task~\cite{sjoberg2003challenges}.
    \item \textbf{Analysis}: The crossover design produces paired data as opposed to the unpaired data of the original experiment, which changes the appropriate hypothesis test~\cite{vegas2015crossover} {(Mann-Whitney U test in the original study vs. Wilcoxon signed-rank test in this study)}. In addition, we extend the original FDA by performing a Bonferroni correction to deal with family-wise error rate when testing multiple hypotheses~\cite{benjamini1995controlling}. Furthermore, we additionally analyze the data using BDA\@. 
\end{itemize}

{Our experiment differs from the original experiment~\mbox{\cite{femmer2014impact}} in all elements~\mbox{\cite{gomez2010replications}}.
However, because a subset of our hypotheses aligns with their hypotheses, part of our study counts as a conceptual replication~\mbox{\cite{baldassarre2014replication}} since it contributes diagnostic evidence for the original claims~\mbox{\cite{nosek2020replication}}.
}
In the rest of this subsection, we report the design of our experiment following the guidelines by Jedlitschka et al.~\cite{jedlitschka2008reporting}.

\subsubsection{Experimental Task}
\label{sec:method:task}

We simulate the use of a requirements specification by subjecting participants to a requirements processing activity, i.e., a common task representing the use of requirements~\cite{femmer2014impact}.
In particular, we present four single-sentence, natural language requirements to the participants and request them to derive a domain model for each of them. 
\Cref{fig:experiment:example} visualizes the expected domain model for the requirement ``Every research object is represented in a JSON-LD format and stored in a document database if it contains a CC license.'' which contains both of the two seeded quality defects.
These defects result in the following challenges according to literature~\cite{pohl2016requirements}:

\begin{enumerate}
    \item The verb in \textbf{passive voice} omits an important entity of the requirement; i.e., that \textit{the data processing unit} stores the research object in a document database (Label 1 in \Cref{fig:experiment:example}).
    \item The \textbf{ambiguous pronoun} ``it'' can syntactically be connected to several preceding noun phrases (``Every research object'', ``JSON-LD format'', and ``a document database'' or the implicit ``Data Processing Unit'') by a reader but semantically only applies to the research object (Label 2 in \Cref{fig:experiment:example}).
\end{enumerate}

\begin{figure}
    \centering
    \includegraphics[width=0.8\textwidth]{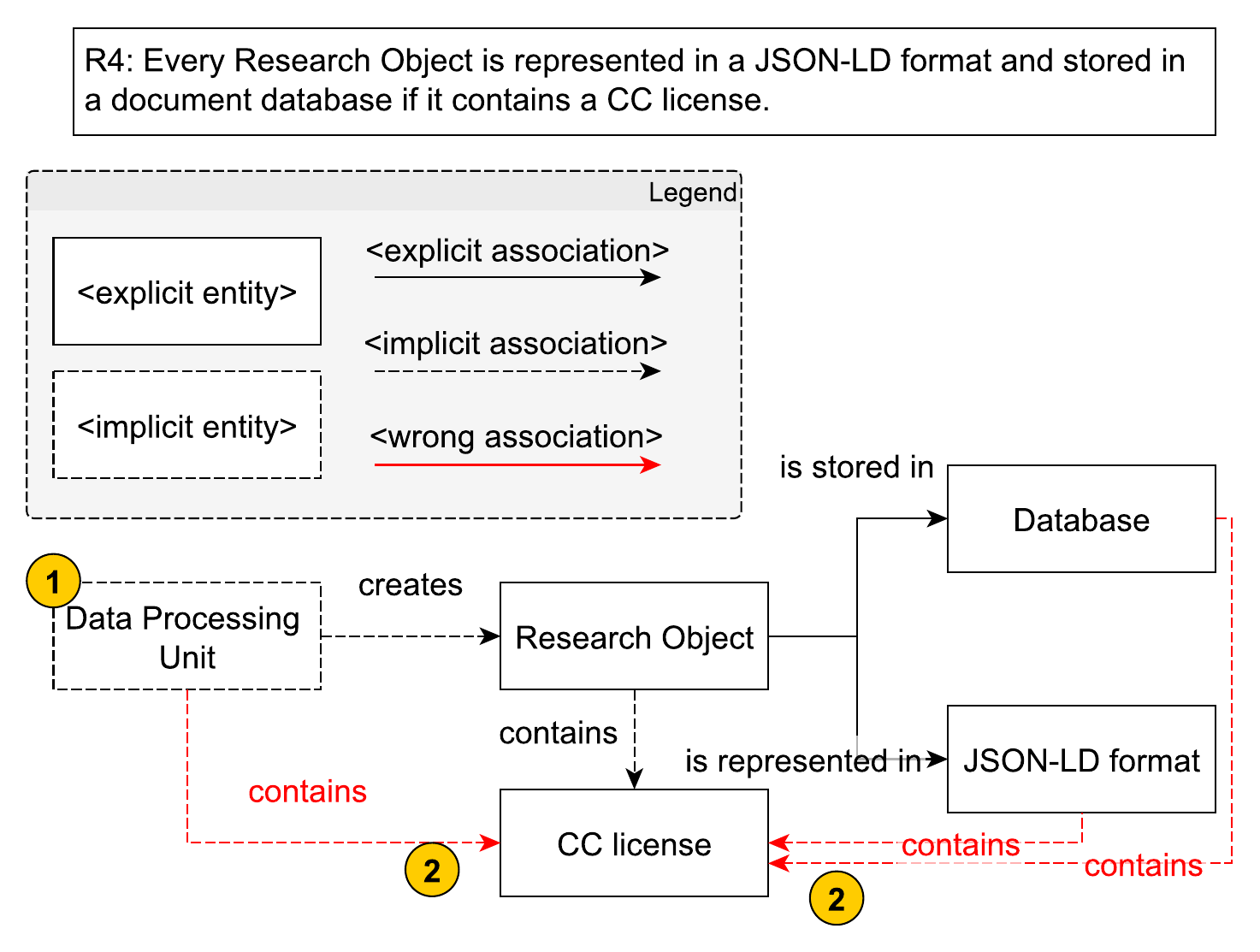}
    \caption{Domain modeling task example for requirement 4.}
    \label{fig:experiment:example}
\end{figure}

The goal of the experimental task is to derive a semantically correct domain model from the natural language requirement which includes identifying all entities (including the implicit ones) and connecting these entities correctly (including those derived from syntactically vague associations). 

The selection of dependent variables was driven by the activity-based requirements quality theory~\cite{femmer2015ABREQM,frattini2023requirements}.
Accordingly, requirements quality is measured by the effect that quality factors have on the relevant attributes of requirements-dependent activity.
We selected the following dependent variables representing the relevant attributes of the domain-modeling activity with the given motivation:

\begin{itemize}
    \item \textbf{Duration}: the longer the domain modeling task takes, the more expensive it is.
    \item \textbf{Number of missing entities}: entities missing from the domain model produce potential cost for failing to involve the respective actor or object.
    \item \textbf{Number of superfluous entities}: entities added to the domain model but not implied by the requirement unnecessarily constrain the solution space.
    \item \textbf{Number of missing associations}: associations missing from the domain model produce a potential cost for failing to identify a dependency between two entities.
    \item \textbf{Number of wrong associations}: associations connecting two entities that establish an unnecessary dependency between them while neglecting an actual dependency.
\end{itemize}

We characterize the domain modeling activity in terms of immediacy (duration), completeness (missing entities and associations), conciseness (superfluous entities), and correctness (wrong associations).

\subsubsection{Hypotheses}
\label{sec:method:hypotheses}


The three independent variables $ind \in \{PV, AP, \text{\textit{PVAP}}\}$ (passive voice, ambiguous pronoun, and the coexistence of passive voice and ambiguous pronoun) and the five dependent variables $dep \in \{D, E^-, E^+, A^-, A^\times\}$ (duration, missing entities, superfluous entities, missing associations, wrong associations) define our 15 null hypotheses as follows.

$$\sum_{ind \in \{PV, AP, \text{\textit{PVAP}}\}} \sum_{dep \in \{D, E^-, E^+, A^-, A^\times\}} H_0^{ind \rightarrow dep}$$

``There is no difference in \textit{\{dep\}} of the domain models based on requirements specifications containing no quality defect and requirements specifications containing \textit{\{ind\}}.''

To capture the context of the experiment, we collected factors based on related work~\cite{petersen2009context,mund2015does,juergens2010much}, including the experience of a practitioner regarding software and requirements engineering, but also in SE roles and in the modeling task itself.
Additionally, we assume that the practitioners' education and domain knowledge influence the dependent variables.
\Cref{tab:vars} summarizes the variables involved in this study.

\begin{sidewaystable}
    \centering
    \caption{Variables of the study (\textbf{ind}ependent, \textbf{con}text, and \textbf{dep}endent).}
    \label{tab:vars}
    \footnotesize
    \begin{tabularx}{\textwidth}{m{2cm} m{1.1cm}|l|X|l|m{5cm}}
        \textbf{Variable} & \textbf{Name} & \textbf{Type} & \textbf{Description} & \textbf{Data type} & \textbf{Range} \\ \hline
        Requirements Quality Defect & RQD & ind & The use of a verb in passive voice, an ambiguous pronoun, or both & categorical & \{none, PV, AP, PVAP\} \\ \hline
        Experience in SE & exp.se & con & Years of experience in software engineering & count &  $\mathbb{N}$ \\
        Experience in RE & exp.re & con & Years of experience in requirements engineering & count &  $\mathbb{N}$ \\
        Education & edu & con & Highest acquired degree & ordinal & \{High School, B.Sc., M.Sc., Ph.D.\} \\
        Primary role & role & con & SE-related Role with the most years of professional experience & categorical & \{requirements engineer, product owner, software architect, developer, tester, quality engineer, trainer, manager, {other}, none\}   \\
        Task experience & exp.task & con & Experience with the task of domain modeling & ordinal & \{never, rarely, from time to time, often\} \\
        Formal modeling training & formal & con & Formal training in domain modeling & categorical & \{true,false\} \\
        Domain knowledge in \{domain\} & dom. \{domain\} & con & Knowledge of the domain $\in$ \{telemetry, aeronautics, databases, open science\} & ordinal & \{1,2,3,4,5\} \\ 
        {Tool experience} & {tool} & {con} & {Experience with using Google Docs for modeling} & {ordinal} & \{{none, rarely, from time to time, often}\} \\
        \hline
        Duration & $D$ & dep & Number of minutes it took the participant to complete the experimental task on one requirement & count & $\mathbb{N}$\\
        Missing entities & $E^-$ & dep & Number of relevant entities missing from the submitted domain model & count & $[0, E_{expected}]$ \\
        Superfluous entities & $E^+$ & dep & Number of not relevant entities included in the submitted domain model & count & $\mathbb{N}$ \\
        Missing associations & $A^-$ & dep & Number of associations missing from the submitted domain model & count & $[0, A_{expected}]$ \\
        Wrong associations & $A^\times$ & dep & Number of associations where either the source or target of the edge is a different entity than implied by the requirement & count & $[0, A_{found}]$ \\
    \end{tabularx}
\end{sidewaystable}

The variables in \Cref{tab:vars} do not include a \textit{participant type} that distinguishes students from practitioners.
While including such a variable is common practice in SE research~\cite{bogner2023restful}, meta-research on the eligibility of students as experiment participants suggests that the labels \textit{student} or \textit{practitioner} are merely a proxy for levels of more meaningful factors like domain knowledge and experience~\cite{salman2015students}.
Additionally, the line between students and practitioners becomes increasingly blurred as students more commonly gather industrial experience before or during their studies~\cite{carver2004issues}.
Consequently, we subsume the participant type variable by the causally more meaningful and fine-grained variables of experience, education, domain knowledge, and formal modeling training.
{We compared two models---one using the binary distinction and one using the more fine-grained variables---and determined that the latter outperforms the former in predictive power, even though only slightly.
This confirms to us that the variables we used are at least as expressive as the binary participant type variable.}

\subsubsection{Experimental Design}
\label{sec:method:design}

Our experimental design includes one factor (RQD) representing the alleged quality defect seeded in a requirements specification.
This main factor contains four treatments: a control one (no defects) and three experimental ones (passive voice (PV), ambiguous pronoun (AP), and both (PVAP)). 

Given our sampling strategy involving industry practitioners, which are difficult to recruit for controlled experiments~\cite{sjoberg2003challenges}, we anticipated a moderate sample size of participants.
Consequently, we opted for a crossover design~\cite{vegas2015crossover} instead of a parallel design, i.e., we apply every treatment to all subjects instead of distributing the subjects among the treatments.
Previously, Kitchenham et al.\ advised against the use of crossover designs~\cite{kitchenham2003case}.
Mainly, the validity of crossover design experiments is challenged by the following confounding factors:

\begin{enumerate}
    \item the period in which a treatment is applied to a subject, as certain periods may influence the dependent variables (e.g., participants may mature and perform increasingly better the more often they perform the experimental task subsequently);
    \item the sequence in which the treatments are applied to a subject, as certain sequences may have a beneficial effect on a dependent variable (e.g., there might be an optimal sequence to apply the treatments in); 
    \item the effect from a previous treatment may \textit{carry over} to the period when applying a subsequent treatment~\cite{vegas2015crossover}; and
    \item the subject variability, as software engineering tasks are highly dependent on the skill of involved individuals~\cite{pickard1998combining}.
\end{enumerate}

However, recent adoptions of best practices from other disciplines made this design applicable to SE research without compromising the validity of the results~\cite{vegas2015crossover,fucci2016an}.
The threats to validity can be mitigated during design and analysis~\cite{vegas2015crossover} by (1) randomizing the order of treatments and (2) including the independent variables' period, sequence, the interaction between them (representing the carryover effect), and subject variability in the analysis.
{Consequently, we consider the \textit{experimental design variables} listed in \mbox{\Cref{tab:vars:exp}} in addition to the variables listed in \mbox{\Cref{tab:vars}}.}

\begin{table}
    \centering
    \caption{{Variables of the study (experimental design factors)}}
    \label{tab:vars:exp}
    \footnotesize
    \begin{tabularx}{\textwidth}{p{1.5cm}|X|l|p{3cm}}
        \textbf{Variable} & \textbf{Description} & \textbf{Data type} & \textbf{Range} \\ 
        \hline
        Period & Index of the experimental period in which the data was obtained & ordinal & [1; 4] \\
        Sequence & Order in which a participants received the treatments & categorical & \{1234, 1243, ..., 4321\} \\
        Carryover effect & Interaction between the period and the treatment & categorical & \{$1\times none$, $1 \times PV$, ..., $4 \times PVAP$ \} \\
        Subject variability & Index of a participant & categorical & \{1, 2, ..., 25\} \\
    \end{tabularx}
\end{table}

When controlling the threats to validity, the crossover design provides two benefits.
Firstly, it requires fewer participants, as an experiment with $n_p$ participants and $n_t$ treatments yields $n_p \times n_t$ observations instead of only $n_p$~\cite{wohlin2012experimentation}.
Secondly, it accounts for subject variability, as the dependent variables can be measured in relation to the average response of each subject instead of the average response of each treatment group~\cite{kitchenham2003case}.
Therefore, each subject acts as its own control and mitigates within-subject variability.

Each experimental session contained four main periods in which we applied one treatment to the subject.
We randomized the order of treatment application to disperse the confounding sequence and carryover effect~\cite{brown1980crossover,vegas2015crossover}.
This resulted in 24 unique sequences of treatment application ($n_t!=24$) and, consequently, 24 experimental groups.
The experiment was single-blinded---i.e., the participants did not know the requirements' sequence, but the researchers did.

\subsubsection{Objects}
\label{sec:method:objects}

The experimental object consisted of four English, single-sentence NL requirements specifications $R_1$-$R_4$. 
An additional warm-up object ($R_0$) preceded the actual experimental objects, adding a fifth experimental period to each session.
It was only used to familiarize the participants with the experimental task and tool and was not considered in the data analysis.
The four experimental objects were manually seeded with defects corresponding to our four treatments: 
one requirement containing none of the two faults, one containing a verb in \textit{passive voice}, one containing an \textit{ambiguous pronoun}, and one containing both a verb in passive voice and an ambiguous pronoun.
The requirements' mean length is 17.8 words (sd=4). 


The first author derived the experimental objects from the requirements specification of the Mars Express ASPERA-3 Processing and Archiving Facility (APAF), a real-world specification from the PuRE data set~\cite{ferrari2017pure}.
From this requirements specification, the first author selected five single-sentence natural language requirements and modified them to ensure two defect-free requirements (one warm-up object $R_0$ and one for the defect-free baseline $R_1$) and three objects with the respective defects ($R_2$-$R_4$).
The second author reviewed and adjusted the selected objects.

\subsubsection{Subjects}
\label{sec:method:subjects}

The target population of interest consists of people involved in software engineering who work with requirements specifications.
We used a non-probability sampling approach based on a mix of purposive and convenience sampling~\cite{baltes2022sampling}.
In particular, we wanted to select participants who are diverse in terms of the context variables as determined in \Cref{tab:vars}, including their experience, education, and software engineering roles.
We approached both students participating in RE courses at our respective institutions and practitioners in our collaborators' network to purposefully diversify the experience and education of our sample.
For all other demographic context factors (e.g., SE roles) we had to rely on convenience sampling.

We approached 52 potential candidates (32 practitioners and 20 students), and 27 candidates (19 \& 8) agreed to participate in the experiment (response rate of 52\%).
Two students did not show up to the agreed time slot.
Our final sample includes 19 practitioners from six different companies and six students from two universities. 
Participation in the experiment was entirely voluntary and not compensated.
The number of participants ($n_p=25$) exceeded the number of experimental sequences ($n_t!=24$) such that we had at least one subject per experimental group and evenly dispersed any confounding sequence or carryover effect~\cite{vegas2015crossover}.

\Cref{fig:demo:sere} shows the distribution of experiment subjects' experience in SE and RE in years. which is widespread in our sample.
Among the 25 participants, the reported primary roles are developer (10), architect (6), requirements engineer (5), manager (1), and no prior professional role (3).
Students who have not yet had a professional software engineering role constitute this last group.
The distribution covers many SE-relevant roles but excludes others, such as testers or product owners.

\begin{figure}
    \centering
    \includegraphics[width=0.8\textwidth]{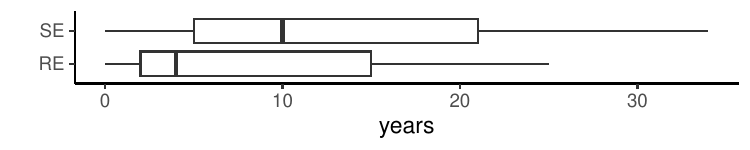}
    \caption{Distribution of SE and RE experience.}
    \label{fig:demo:sere}
\end{figure}

Among the $25$ participants, $5$ reported a high school degree as their level of education, $8$ a Bachelor's degree, and $12$ a Master's degree. 
No participant reported a Ph.D.\ degree as their highest degree of education.
\Cref{fig:demo:dom} visualizes the distribution of the participant's experience in the four domains\footnote{{Domain does not exclusively mean application domain, but rather any coherent ontology related to a specific topic.}} that {contribute semantic knowledge to understanding the requirements}: aeronautics, telemetry, databases, and open source.
For the latter two, results are balanced across the different knowledge levels, while the former two confirm our assumption that all participants had a low-level knowledge of aeronautics and telemetry systems.

\begin{figure}
    \centering
    \includegraphics[width=0.8\textwidth]{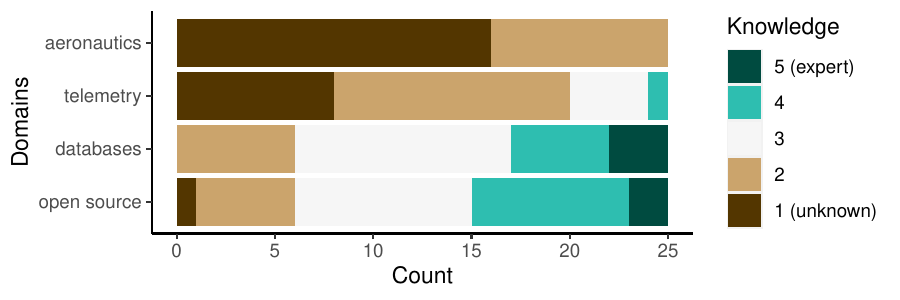}
    \caption{Distribution of knowledge in the four relevant domains.}
    \label{fig:demo:dom}
\end{figure}

The experiment tool---the Google Draw plugin within the Google Document---was unknown to most ($18$ never used it, $6$ rarely, and $1$ from time to time).
A total of $16$ participants ($64$\%) report having received a form of training in the modeling activity. 
The modeling experience (never: $1$, rarely: $11$, from time to time: $10$, often: $3$) resembles a normal distribution.
{We did not discard data from participants who reported having no modeling experience or formal training in modeling given that our experiment included both comprehensive instructions and a warm-up phase as described in \mbox{\Cref{sec:method:objects}}.
}

Given the distribution of responses in these context variables, we disqualified the following predictors:
aeronautics domain knowledge, telemetry domain knowledge, and experience with the experiment tool.
These variables are not sufficiently distributed in our sample of study participants, i.e., several categories of these variables are underrepresented.
Consequently, they are unable to effectively block the influence of that variable on the dependent variable~\cite{wohlin2012experimentation}.

\subsubsection{Instrumentation}
\label{sec:method:instrument}

We used a Google Docs document\footnote{\url{https://www.google.de/intl/en/docs/about/}} for the task and a Google Form questionnaire\footnote{\url{https://www.google.com/forms/about/}} to collect demographic information.
The Google Docs document lent itself to the task due to its accessibility and its simple modeling tool with the embedded Google Drawings.
The modeling tool represented the optimal trade-off between complexity---as neither previous knowledge nor additional software was necessary to conduct the experimental task---and suitability---as it contains all elements relevant to the domain modeling task (i.e., nodes for entities and edges for associations).
This main study document explained the experimental task, an example of the domain modeling task, and a short context description of the system from its original requirements specification~\cite{ferrari2017pure}.

We created a survey questionnaire to collect demographic information relevant to the experiment using Google Forms.
All participants could answer the survey only after completing the task to avoid fatiguing effects.
At the beginning of the questionnaire, participants entered their assigned participant ID (PID), such that we could connect their response to the experimental task to their response to the questionnaire without storing any personal data. 
The questions were designed to collect all relevant independent variables listed in \Cref{tab:vars}.

We piloted the experiment in a session with two Ph.D.\ students in SE\@. 
We clarified the instruction text and task descriptions based on the collected feedback.

\subsubsection{Data Collection Procedure}
\label{sec:method:procedure}

We scheduled a one-hour session according to the availability of the participants.
Because of differing schedules and time zones, we scheduled 16 sessions with up to three participants simultaneously. 
We conducted the sessions between 2023-04-03 and 2023-04-17.

Each session started with the first author explaining the general procedure of the experiment and obtaining consent to evaluate and disclose the anonymized data.
No participant refused this consent and all data points could be included in the data evaluation procedure.
Then, participants were instructed to read the prepared document in order and complete the contained tasks.
The document contained all descriptions of the task such that all participants received the same instructions.
The first author oversaw all sessions to address technical difficulties and recorded the minutes each participant spent per period.
Ten minutes were estimated per period, but participants were free to allocate their time.
In case participants took longer than the scheduled one hour, they completed the task in as much time as they required.
Once the task was complete, participants also filled in the questionnaire to provide demographic information on context variables.

\subsubsection{Data Preparation}
\label{sec:method:prep}

To evaluate the collected data, we created a code book that characterizes issues in domain models.
We developed detection rules for each dependent variable of the resulting product---i.e., missing entity, superfluous entity, missing association, and wrong association---{and summarized them in a guideline (available in our replication package~\mbox{\cite{frattini2024replication}}).}
Then, the first author manually evaluated the resulting domain models of each participant using this guideline and recorded all detected issues.

The result of the coding process was a table where one row represents the evaluation of one domain model.
Given $n_p=25$ participants and $n_r=4$ requirements, we ended up with $n_p \times n_r=100$ data points.
Each data point contained the number of issues of each of the four types that occurred in the respective domain model.
Finally, we standardized numerical variables in the demographic data for easier processing.

To assess the reliability of the rating, the fourth author of the paper independently recorded issues of three randomly selected participant responses, yielding an overlap of twelve ratings.
Since each domain model can contain an arbitrary number of issues of each type, but our dependent variables only model the number of times that an issue type occurred, we consider each rating of a domain model as a vector of dimensions equal to the number of issue types. 
{We then calculated the inter-rater agreement of the same domain model using the Spearman rank correlation between the vectors.
The average cosine similarity is 77.0\% and represents substantial agreement.
The two raters discussed the remaining disagreement and concluded that they represented acceptable variance in the interpretation of participants' responses.}

\subsubsection{Frequentist Data Analysis}
\label{sec:method:analysis:fda}

We performed a frequentist data analysis of the experimental data as in the original experiment~\cite{femmer2014impact}. 
Since the factor is categorical, all dependent variables are continuous, and our samples are dependent, our statistical method of choice falls between the parametric paired t-test~\cite{hsu2014paired} or the non-parametric {Wilcoxon signed-rank test}~\cite{wilcoxon1945individual} based on the distribution of the variables, which we evaluated using the Shapiro-Wilk test results~\cite{shapiro1965analysis}.

We reject a null hypothesis if the resulting p-value of a {two-tailed} statistical test is lower than the significance level $\alpha$. 
To account for type I errors when performing multiple hypotheses tests targeting the same independent variable, we apply the Bonferroni correction~\cite{benjamini1995controlling}: 
we considered $\alpha' = \frac{\alpha}{m}$ where $\alpha = 0.05$ and $m$ is the number of hypotheses tested for each value of the independent variable.
For our five families of hypotheses $\alpha' = \frac{0.05}{5} = 0.01$. 

Additionally, we report the effect size~\cite{dybaa2006systematic} using Cohen's D for the paired student t-test~\cite{cohen1969statistical} and the matched-pairs rank biserial correlation coefficient for {Wilcoxon signed-rank test}~\cite{king2018statistical}.

\subsubsection{Bayesian Data Analysis}
\label{sec:method:analysis:bda}

We apply Bayesian data analysis with Pearl's framework for causal inference~\cite{pearl2016causal} to complement the frequentist data analysis~\cite{furia2019bayesian}.
Given the limited adoption of Bayesian data analysis in software engineering research~\cite{furia2019bayesian, furia2022applying, furia2023towards}, we complement this method section with a running example for understandability.
In this running example, we illustrate the methodological steps of Bayesian data analysis for the hypothesis that requirements quality defects influence the number of wrong associations in a resulting domain model.

Three steps~\cite{siebert2023applications}, which are the major steps of Pearl's original model of causal statistical inference~\cite{pearl2016causal}, comprise the analysis. 
We explain each step in the following paragraphs.

\paragraph{Modeling}

In the modeling step, we make our causal assumptions about the underlying effect explicit in a graphical causal model.
The graphical causal model takes the form of a directed acyclic graph (DAG) in which nodes represent variables and directed edges represent causal effects~\cite{elwert2013graphical}.
Our DAG contains four groups of variables:

\begin{enumerate}
    \item Treatment: The independent variable that represents the requirements quality defect present in the requirement.
    \item Context factors: The independent variables that represent the properties of the participants.
    \item {Experimental design factors: The independent variables that represent all factors of the crossover experiment design influencing the response variables~\mbox{\cite{vegas2015crossover}}.}
    \item Response variables: The dependent variables.
\end{enumerate}

The effect of the treatments on the response variables is the subject of the analysis.
By including both context and confounding factors, their influence is factored out from the treatments' causal effect on the response variables.
Consequently, the effect of interest can be isolated from any confounding factor included in the DAG.
We assume causal relations---represented by edges in the DAG---between every independent (treatment, context, and confounding) and the dependent variables.
Additionally, independent variables may influence other independent variables.

\Cref{fig:dag:wrongassociations} shows the DAG of the running example.
The treatment, response variable, and context factors correspond to the study variables as outlined in \Cref{tab:vars}.
{The experimental design variables correspond to the factors listed in \mbox{\Cref{tab:vars:exp}}.}
The \textit{experimental period} blocks the learning effect, i.e., the influence on the response variable caused by repeatedly performing the task.
The \textit{duration} blocks the time effect, i.e., the influence on the response variable caused by the amount of time that a participant took for each instance of the task.
Note that the factor \textit{tool experience} listed in \Cref{tab:vars} is missing from the DAG since we excluded it as explained in \Cref{sec:method:subjects}.
Note that the factor \textit{sequence} listed in \Cref{tab:vars:exp} is missing from the DAG since it is confounded with subject variability (further explained in \Cref{sec:discussion:threats:internal}).

\begin{figure}
    \centering
    \includegraphics[width=0.9\textwidth]{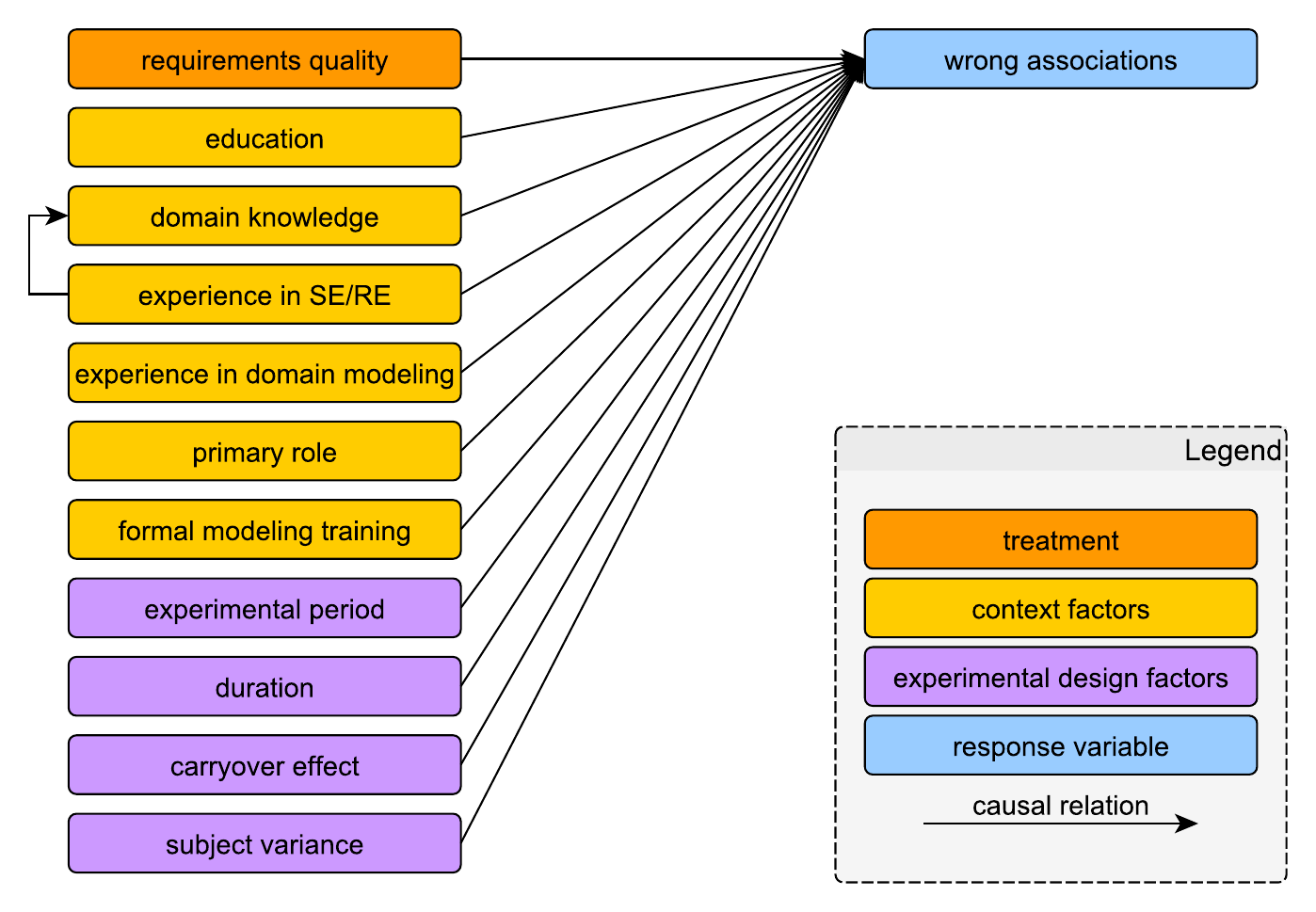}
    \caption{DAG for the analysis of wrong associations}
    \label{fig:dag:wrongassociations}
\end{figure}

The DAG does not visualize the \textit{interaction effects} we assume between two independent variables. 
An interaction effect occurs when the influence of one independent variable on the dependent variable depends on the value of another independent variable~\cite{mcelreath2020statistical}.
Visualizations of interaction effects in DAGs have been proposed~\cite{nilsson2021directed} but are not common practice.
In the running example, we assume two interaction effects via the following hypotheses:

\begin{enumerate}
    \item \textit{requirements quality * domain knowledge}: domain knowledge can compensate the effect of requirements quality defects~\cite{poesio1996ambiguity}
    \item \textit{requirements quality * period} carryover effect~\cite{vegas2015crossover}: the effect of a treatment may be influenced by the treatments applied in previous periods
\end{enumerate}

\paragraph{Identification}

{Including an independent variable Z that has an assumed causal effect on both the treatment X (i.e., $Z \rightarrow X$) and the outcome Y (i.e., $Z \rightarrow Y$) opens a non-causal path (i.e., $X \leftarrow Z \rightarrow Y$) from the treatment to the outcome~\mbox{\cite{pearl1995bayesian}}.
This so-called backdoor path introduces spurious associations.}
Consequently, blindly moving forward with all variables may harm the causal analysis.
Instead, the so-called adjustment set of variables needs to be selected~\cite{mcelreath2020statistical} in the \textit{identification} step.
A series of four criteria~\cite{mcelreath2020statistical} allows to make an informed selection of variables to include in the final estimation step.
This way, we avoid variable bias like colliders which confound the causal effect between the treatment and the response variable.

In the running example, we assume the following causal relation between independent variables.
The more experience a participant has in SE or RE, the more likely it is that they have acquired respective domain knowledge (experience in SE\slash RE $\rightarrow$ domain knowledge).
We need to consider this relationship in the next step to avoid attributing impact to the wrong independent variable.
For instance, in the running example, we need to distinguish whether \textit{experience in SE\slash RE} has a direct influence on \textit{wrong association} or whether it just influences \textit{domain knowledge}, which influences the response variable. 

Because we employ an experiment as our research method and fully control the treatment, there is no influence of any other independent variable on the treatment variable.

\paragraph{Estimation}

In the estimation step, we perform a regression analysis.
The regression analysis results in estimates of the response variable depending on the values of the independent variables.
The result of the regression analysis is a Bayesian model trained with empirical data.
The model provides the magnitude and sign of the effect that each independent variable has on the dependent response variable.

The estimation step begins by selecting a distribution type (likelihood) that represents the dependent response variable~\cite{gelman2020bayesian}.
We select the distribution type based on the maximum entropy criterion~\cite{jaynes03} and ontological assumptions.
This means we select the least restrictive distribution that fulfills all ontological assumptions about the variables' properties.

In our running example, the response variable is a count of wrong associations in a domain model.
Consequently, the distribution must be discrete and only allow positive numbers or zero.
Additionally, the response variable is bounded by the number of \textit{expected associations} of the domain model, i.e., the number of associations in the sample solution, since a participant can only connect as many associations wrongly in the model as there were associations expected.
Any associations added beyond the expected associations count as \textit{superfluous associations}, a different response variable.
Consequently, we represent the response variable with a \textit{Binomial} distribution.
The following formula encodes that the number of wrong associations in one domain model i ($E_i^\times$) is distributed as a Binomial distribution with the number of trials equal to the number of expected associations ($E$) and a probability $p_i \in [0, 1]$ of getting one association wrong.

$$ E_i^\times \sim Binomial(E, p_i) $$

This formula assumes that the event---connecting one association wrong---is independent, i.e., one wrong association does not influence the success of any other association.

In the next step, we define the parameter that determines the response variable distribution (in the running example: $p_i$) in relation to the predictors selected in the identification step.
The following formula shows a simplified version of this parameter definition (excluding most of the previously mentioned predictors in \Cref{tab:vars} for brevity).

$$ logit(p_i) = \alpha + \alpha_{PID} + \beta_{RQD}^T \times RQD_i + \beta_{SE} \times exp.se_i $$

The \texttt{logit} operator scales the parameter $p_i$ to a range of $[0, 1]$ since the probability parameter of the Binomial distribution only accepts this range of values~\cite{demaris1992logit,mcelreath2020statistical}.
The parameter $p_i$ is, in this example, determined by the following predictors:

\begin{enumerate}
    \item Intercept ($\alpha$): the grand mean of connecting an association wrongly, i.e., the baseline challenge of getting an association wrong.
    \item Group-level intercept ($\alpha_{PID}$, where the results of one participant represent one group): the participant-specific mean of connecting an association wrongly, i.e., the within-subject variability of response variables~\cite{vegas2015crossover} modeled via partial pooling~\cite{ernst2018bayesian}
    \item Treatment ($RQD_i$): the influence of a requirements quality defect on the probability of connecting an association wrong (as an offset from the grand mean).
    \item Software Engineering Experience ($exp.se_i$): the influence of the subject's software engineering experience on the probability of connecting an association wrong (as an offset from the grand mean).
\end{enumerate}

The variables $RQD_i$ and $exp.se_i$ contain the values recorded during instance $i$ of conducting the experimental task and are each prefixed with coefficients $\beta_{RQD}^T$ and $\beta_{SE}$.
These coefficients are Gaussian probability distributions that represent the magnitude and direction of the influence that the variable values have on the parameter $p_i$ and, therefore, on the distribution of the response variable.
The mean $\mu$ of the coefficient represents the average effect of the variable on the parameter $p_i$, and the standard deviation $\sigma$ encodes the variation around this average effect.
A standard deviation of $\sigma=0$ would mean that the variable has a deterministic effect of strength $\mu$ on $p_i$ and, therefore, the distribution of the response variable.
In reality, this is highly unrealistic.
Hence, the standard deviation captures the uncertainty of the effect of a variable on $p_i$.

Special cases of variables are the intercepts $\alpha$ and $\alpha_{PID}$, which are probability distributions without any variable and, hence, represent the predictor-independent general and participant-specific probability of connecting an association wrong.
In the beginning, we assign probability distributions spread around $\mu=0$ ($\beta_{RQD}^T \sim Normal(0, 0.5)$), so-called \textit{uninformative priors}, to these coefficients.
These distributions encode our prior beliefs about the influence of the respective predictor, i.e., that it is yet unknown whether the predictor has a positive ($\mu > 0$) or negative ($\mu < 0$) influence on $p_i$.
Only where previous evidence for the impact of a predictor on the response variable exists, we select more informative priors.
For example, the experiment by Femmer et al.~\cite{femmer2014impact} indicates that \textit{missing entities} and \textit{missing associations} are in general rare, which we represent in our priors by selecting $\alpha \sim Normal(-1, 0.5)$ for the intercept.

We assess the feasibility of the selected prior distributions via prior predictive checks~\cite{wesner2021choosing}.
During this check, we sample only from the priors, i.e., we predict the response variable given the recorded data of independent variables and the prior probability distributions of the predictor coefficients. 
\Cref{fig:checks} visualizes the result of the prior predictive check.
The grey bars represent the actual observed distribution of the response variable.
For example, 75 domain models contained zero wrong associations.
The distribution of predicted values for the response variable (cyan whisker plots) encompasses the actual observed distribution of the response variable.
This confirms that the actually observed distribution is approximately determined by the uninformative prior distributions.

\begin{figure}
    \centering
    \includegraphics[width=\textwidth]{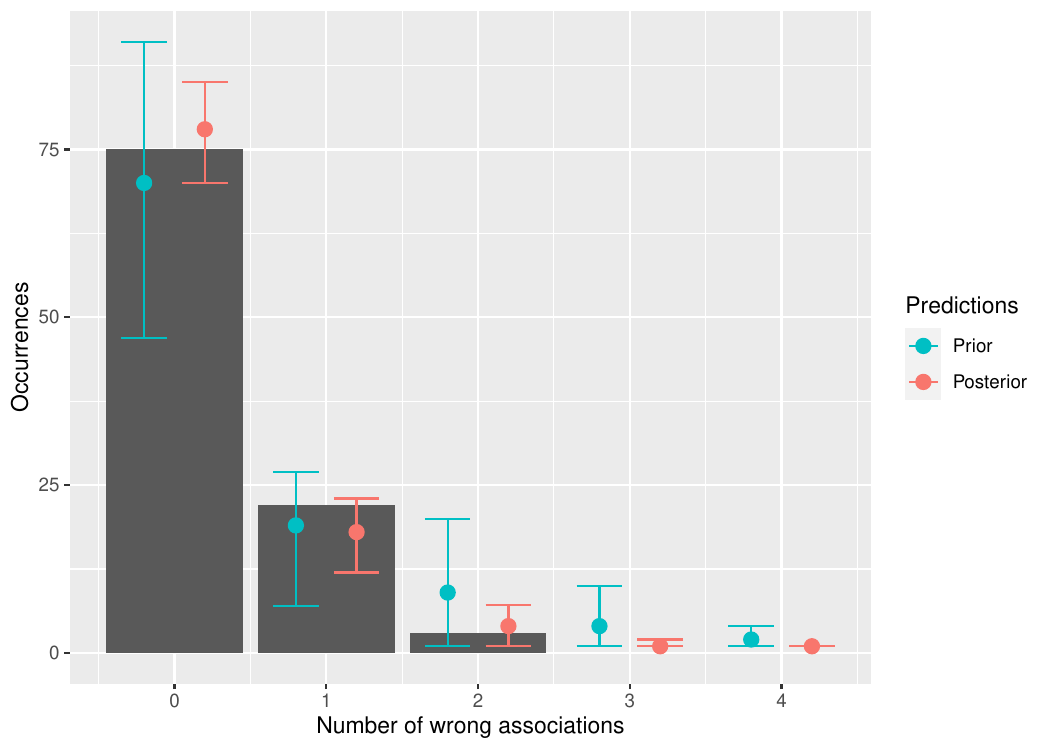}
    \caption{Predictive checks with prior and updated posterior coefficient distributions}
    \label{fig:checks}
\end{figure}

Upon confirmation of the priors' feasibility, we train the Bayesian models with the data recorded during the experiment.
We conducted the analysis using the \texttt{brms} library~\cite{burkner2017brms} in \textsl{R}.
Hamiltonian Monte Carlo Markov Chains (MCMC)~\cite{brooks2011handbook} update the coefficient distributions based on the empirical data.
During this process, the parameters of the coefficient distributions are adjusted to better reflect the response variable based on the predictor variable values.

After the training process, we perform posterior predictive checks, which work similarly to the prior predictive check but use the updated posterior coefficient distributions instead of the prior distributions. 
\Cref{fig:checks} also visualizes the posterior predictive check for the running example.
The distribution of the predicted values (red whisker plots) still encompasses the actually observed distribution of the response variable but has narrowed around these values.
This indicates that the posterior distributions encode the influence of the predictor variables more accurately than the prior distributions, i.e., that the model has successfully gained predictive power during the training process.

To overcome the problem mentioned in the identification step, i.e., attributing impact to the wrong predictor, we train additional models per response variable to test for conditional independence~\cite{mcelreath2020statistical}.
For example, to determine the correct causal relationship between the two independent variables \textit{experience in SE}, \textit{domain knowledge}, and the dependent variable, we train two additional models where each one misses one of the two variables~\cite{furia2023towards}.
After training, we compare the posterior distributions of the remaining parameter coefficients.
If a posterior distribution significantly moves from $|\mu|>0$ towards $\mu=0$ when including a variable, then the response variable is independent of that variable when conditioning on the included variable.
The model does not gain any further information from the variable with $\mu \simeq 0$, and its causal relation is disputed.
If the posterior distribution does not deviate significantly when including another variable, its causal impact is confirmed.

Finally, we perform a stratified posterior prediction to answer our research questions.
To this end, we construct a synthetic data set with four data points, one for each value of the main factor variable (i.e., baseline, PV, AP, PVAP).
We fix all other independent variables at representative values---i.e., the mean for continuous and the mode for discrete variables. 
Then, we sampled $6,000$ predictions for each of the four data points.
This isolates the effect of the treatment but maintains the uncertainty of the influence of every independent variable encoded in the standard deviation of every predictor coefficient and, hence, more accurately describes the causal relationship between the treatment and the outcome.
We compare the $6,000$ predictions of each of the three treatments (PV, AP, PVAP) with the $6,000$ predictions from the baseline (no defect) and count how often the treatment causes a higher, equal, or lower outcome variable.
We scale these values to percentages to summarize the effect of the treatment on the outcome variable.
This evaluation avoids a point-wise reduction of the results and comparison to an arbitrary significance level as customary in frequentist analyses~\cite{gren2021possible}.
{Rather than providing a binary answer to the hypotheses, we present the more informative distribution of results.
However, for the sake of reporting, we consider the distribution of the duration variable \textit{skewed} if the two percentages differ from the mean (50\%) by 10\% each and consider the other distributions skewed if the two percentages differ by 10\% from each other.
}

Additionally, we plot the marginal effect of selected independent variables to visualize their isolated impact on the response variable.
The isolated impact reveals how context and confounding factors influence the response variable.
This includes visualizing the carryover effect, i.e., the interaction between the treatment and the period.

\section{Results}
\label{sec:results}

\Cref{sec:results:analysis:fda} shows the results of the frequentist and \Cref{sec:results:analysis:bda} the results of the Bayesian data analysis. 
{\mbox{\Cref{sec:result:comparison:replication}} compares the part of our results that contributes a conceptual replication to the original study.
\mbox{\Cref{sec:result:comparison:bda}} compares the results from our FDA to the results from our BDA.
}

\subsection{Frequentist Data Analysis}
\label{sec:results:analysis:fda}

{\mbox{\Cref{tab:results}} shows the mean and median values of the response variables similar to how they are reported by Femmer et al.~\mbox{\cite{femmer2014impact}}.
Note that the results are not directly comparable to those in the original study as both our experimental objects and treatments varied.
}

\begin{table}[hbt!]
    \centering
    \caption{{Mean and median response variable values (reported as mean/median in each cell)}}
    \label{tab:results}
    \begin{tabularx}{\textwidth}{l|XXXXX}
        \toprule
        \textbf{Defect} & \textbf{Duration $D$} & \textbf{Missing Entities $E^-$} & \textbf{Superfluous Entities $E^+$} & \textbf{Missing Associations $A^-$} & \textbf{Wrong Associations $A^\times$} \\
        \midrule
        none & 7.38/6.5 & 0.23/0 & 0.5/0 & 0.38/0 & 0.08/0 \\
        PV & 6.88/7 & 0.81/1 & 0.42/0 & 0.81/1 & 0.62/0 \\
        AP & 7.12/6 & 1.23/1 & 0.96/0.5 & 1.12/1 & 0.54/0.5 \\
        PVAP & 7.96/7 & 1.27/1 & 0.46/0 & 1.38/1 & 0.38/0 \\
        \bottomrule
    \end{tabularx}
\end{table}

\Cref{tab:results:fda} lists the results of our frequentist data analysis and relates them to the results from the original study~\cite{femmer2014impact}.

\begin{table*}[hbt!]
    \centering
    \caption{Results of frequentist analysis including the p-value of the hypothesis test (p), confidence interval (CI), and effect size (ES). Statistically significant results in \textbf{bold} (original study $\alpha=0.05$, this experiment $\alpha' = 0.01$).}
    \label{tab:results:fda}
    \begin{tabular}{p{1.5cm}|l|lrr|lrrr}
        \toprule
        \textbf{Outcome} & \textbf{Treatment} & \multicolumn{3}{c|}{\textbf{Original}~\cite{femmer2014impact}} & \multicolumn{3}{c}{\textbf{Replication}} \\
        & & $p$ & CI & ES & $p$ & CI & ES \\ \midrule 

        \multirow{3}{*}{Duration} & PV & & & & 0.67 & $(-0.4, 0.7)$ & -0.13 \\
        & AP & & & & 0.86 & $(-0.7, 0.86)$ & 0.01 \\
        & PVAP & & & & 0.49 & $(-0.5, 0.25)$ & 0.14 \\ \hline 
        
        Missing Actors & PV & 0.10 & $(0, \infty)$ & 0.39 \\ \hline
        {Missing Objects} & PV & 0.25 & $(-1, \infty)$ & 0.25 \\ 
        \hline
        \multirow{3}{*}{\shortstack[l]{Missing\\Entities}} & PV & & & & $\ll$ \textbf{0.01} & $(-1, 0)$ & -0.79 \\
        & AP & & & & $\ll$ \textbf{0.01} & $(-1.5, 0)$ & -0.93 \\
        & PVAP & & & & $\ll$ \textbf{0.01} & $(-2, 0)$ & -0.81 \\ \hline 
        \multirow{3}{*}{\shortstack[l]{Superfluous\\Entities}} & PV & & & & 0.64 & $(-0.5, 0)$ & 0.14 \\
        & AP & & & & 0.19 & $(-2.5, 0)$ & -0.41 \\
        & PVAP & & & & 0.62 & $(-1, 1)$ & 0.15 \\ \hline 
         
        \multirow{3}{*}{\shortstack[l]{Missing\\Associations}} & PV & \textbf{0.02} & $(1, \infty)$ & 0.75 & 0.025 & $(-1, 0)$ & -0.58 \\
        & AP & & & & $\ll$ \textbf{0.01} & $(-2, -1.5)$ & -0.87 \\
        & PVAP & & & & $\ll$ \textbf{0.01} & $(-2, -1)$ & -0.84 \\ \hline 
        \multirow{3}{*}{\shortstack[l]{Wrong\\Associations}} & PV & & & & 1.0 & $(0, 0)$ & 0.0 \\
        & AP & & & & $\ll$ \textbf{0.01} & $(-1, 0)$ & -0.85 \\
        & PVAP & & & & 0.052 & $(-1.5, 0)$ & -0.67 \\
        \bottomrule
    \end{tabular}
\end{table*}

The frequentist data analysis suggests rejecting the following hypotheses and, therefore, proposes the following effects as statistically significant (with $\alpha' = 0.01$):

\begin{enumerate}
    \item $H_0^{PV \rightarrow E^-}$: passive voice impacts the number of missing entities
    \item $H_0^{AP \rightarrow E^-}$: ambiguous pronouns impacts the number of missing entities
    \item $H_0^{\text{\textit{PVAP}} \rightarrow E^-}$: the co-occurrence of passive voice and ambiguous pronouns impacts the number of missing entities
    \item $H_0^{AP \rightarrow A^-}$: ambiguous pronouns impact the number of missing associations
    \item $H_0^{\text{\textit{PVAP}} \rightarrow A^-}$: the co-occurrence of passive voice and ambiguous pronouns impacts the number of missing associations
    \item $H_0^{AP \rightarrow A^\times}$: ambiguous pronouns impact the number of wrong associations
\end{enumerate}

The associated effect size is considered large~\cite{cohen2008explaining} in all cases.

\subsection{Bayesian Data Analysis}
\label{sec:results:analysis:bda}

This section follows the methodology described in \Cref{sec:method:analysis:bda} by presenting the DAG in \Cref{sec:results:analysis:bda:dag}, posterior predictions in \Cref{sec:results:analysis:bda:posterior}, and marginal plots in \Cref{sec:results:analysis:bda:marginal}.

\subsubsection{Causal Model and Adjustment Set}
\label{sec:results:analysis:bda:dag}

\Cref{fig:dag} visualizes the DAG that graphically models our causal assumptions.
It is an extension of \Cref{fig:dag:wrongassociations}, the running example, including all five dependent variables.
To preserve the readability of the DAG, we introduce a \textit{distributor} node.
This node substitutes the connections from the source of every incoming edge to the target of every outgoing edge.

\begin{figure}
    \centering
    \includegraphics[width=1\textwidth]{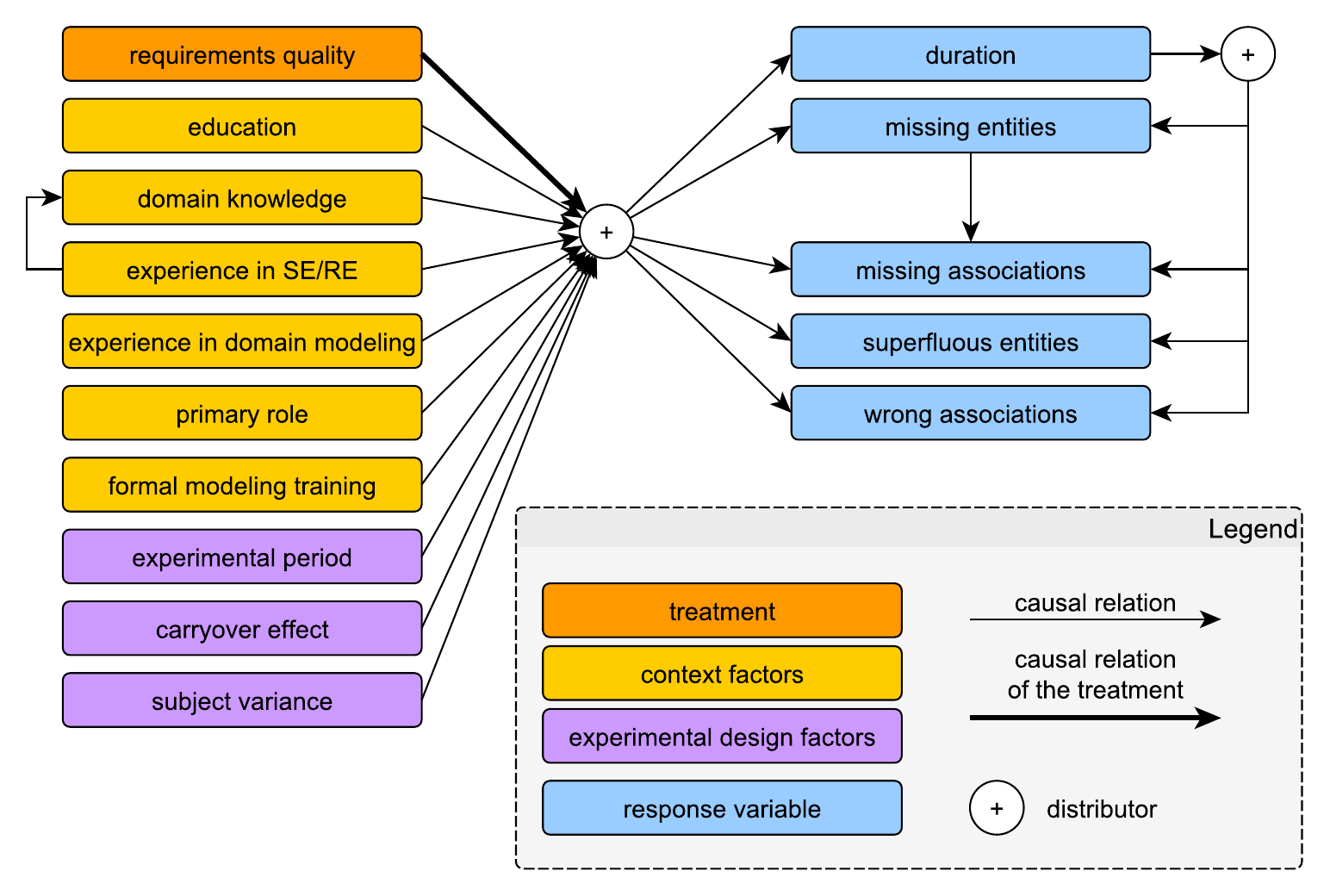}
    \caption{Directed acyclic graph visualizing all causal assumptions}
    \label{fig:dag}
\end{figure}

The edges represent the same causal reasoning as presented in the running example in \Cref{sec:method:analysis:bda}.
We assume that all independent variables (treatments, context factors, and {experimental design factors}) have an impact on all dependent variables.
The impact of the requirements quality defect (the three treatments) is the relationship of interest in our analyses. 
In addition to the already described impact of experience in SE on domain knowledge, we assume the following causal relationships:

\begin{enumerate}
    \item Duration impacts all other dependent variables: Since we did not constrain the time for each period, different amounts of minutes taken for each object may influence the results. Taking a longer time for one domain model may reduce the amount of defects in the final model.
    \item Missing entities impact missing associations: If an entity is missing from the domain model, any association involving that entity will also be missing {(as already supported by our re-analysis~\mbox{\cite{frattini2024second}}).}
\end{enumerate}

{Equally notable are the non-existing associations between nodes, especially between context factors.
In our DAG, we only assume an impact of experience in SE/RE on domain knowledge (as explained in \mbox{\Cref{sec:method:analysis:bda}}).
We do not assume, for example, a causal relation between education and experience in SE/RE as higher levels of education do not entail more industrial experience or vice versa.
Similarly, we assume education to be independent of domain knowledge as most educational programs known to us are domain-independent.
The resulting set of associations visualized in \mbox{\Cref{fig:dag}} corresponds to the authors' shared beliefs that warrant assuming causal relationships between two variables.
While we do not expect every reader to share these beliefs, we hope that the explicit and transparent documentation of our assumptions invites constructive, iterative improvements by challenging them via empirical investigations.
}

The DAG from the running example in \Cref{fig:dag:wrongassociations} lists the variable duration as an independent variable, while the final DAG in \Cref{fig:dag} lists it as a response variable.
The variable duration takes on two distinct roles depending on the current analysis. 
In the case where the analysis targets the effects on the duration, it is the sole response variable.
In all other cases, it is an independent variable.

In the second step in the statistical causal inference~\cite{siebert2023applications}, the identification step, we found the adjustment set to include all variables for prediction.
To discern the impact of independent variables with causal relations among them, we developed comparison models in which certain variables were excluded.
The comparison showed that the exclusion did not change the estimations of the coefficient distributions.
Hence, we assume all causal relationships are feasible and evaluate the full model, including all eligible variables as predictors.

\subsubsection{Posterior Predictions}
\label{sec:results:analysis:bda:posterior}

Based on maximum entropy~\cite{jaynes03}, we model the five response variables using the following probability distributions.
The duration is centered around the global mean, therefore, we model it with a Gaussian distribution around $\mu=0$.
The number of superfluous entities is an unbounded count with an index of dispersion of about 1.5, hence, we model it as a negative binomial distribution.
Missing entities, missing associations, and wrong associations are bounded counts and, hence, modeled as Binomial distributions.
\Cref{tab:posteriors} contains the result of the predictions from the posterior distributions.
Each cell contains the resulting likelihood that the occurrence of a factor causes fewer or more issues of the respective outcome compared to the baseline of no quality defects\footnote{The remaining cases ($100\%-less-more$) are omitted from the table}.
The larger the difference between the likelihood of more ($+$) than fewer ($-$) issues, the stronger the effect of that factor on the outcome.
If the likelihood of more ($+$) issues outweighs the likelihood of fewer ($-$) issues, the factor has a negative effect.
If the two values are similar, then the factor has no clear effect on the outcome.

\begin{table}[ht]
    \centering
    \caption{Likelihood that a treatment produces fewer ($-$) or more ($+$) occurrences of the respective outcome variable.}
    \label{tab:posteriors}
    \begin{tabular}{l|cc|cc|cc}
         \multirow{2}{*}{\textbf{Outcome}} & \multicolumn{2}{c|}{\textbf{PV}} & \multicolumn{2}{c|}{\textbf{AP}} & \multicolumn{2}{c}{\textbf{PVAP}} \\
         & - & + & - & + & - & + \\ \hline
         Duration & 57.2\% & 42.8\% & 51.7\% & 48.3\% & 44.8\% & 55.2\% \\[0.1cm] \hline
         Missing entities & 29.6\% & 32.5\% & 24.7\% & 40.1\% & 27.4\% & 35.6\% \\[0.1cm]
         Superfluous entities & 26.6\% & 22.5\% & 22.5\% & 33.3\% & 26.4\% & 24.6\% \\[0.1cm] \hline
         Missing associations & 25.0\% & 45.0\% & 20.2\% & 51.2\% & 22.2\% & 49.4\% \\[0.1cm]
         Wrong associations & 10.5\% & 11.5\% & 5.2\% & 44.6\% & 6.9\% & 31.5\% \\
    \end{tabular}
\end{table}

For example, the first two cells in the second row of \Cref{tab:posteriors} state that using passive voice causes \textit{fewer} missing entities in $29.6\%$ and \textit{more} missing entities in $32.5\%$ of all cases.
In the remaining $37.9\%$ of all cases, passive voice causes neither more nor fewer missing entities.
Given this balance, the effect of passive voice on missing entities is unclear, and there is not enough evidence to reject $H_0^{PV \rightarrow E^-}$. 

\begin{highlightbox}{Result of Posterior Predictions}
    The following effects are likely given the skewed distribution of posterior predictions: 
    passive voice, ambiguous pronouns, and their co-occurrence cause an increasing number of missing associations. 
    {Ambiguous pronouns cause an increasing number of wrong associations.
    Ambiguous pronouns cause an increasing number of missing and superfluous.}
\end{highlightbox}

{We use an arbitrary threshold of 10\% to report notable results in textual form.
Refer to \mbox{\Cref{tab:posteriors}} for the actual, more fine-grained results.
}


\subsubsection{Marginal and Conditional Effects}
\label{sec:results:analysis:bda:marginal}

Marginal plots visualize the isolated effect of specific predictors when fixing all other predictors to representative values.
In the following, we present selected marginal plots that show the effects of interest.
The remaining plots can be found in our replication package.

\paragraph{Missing entities impact missing associations}

\Cref{fig:missingentities} visualizes the effect of the number of missing entities on the number of missing associations.
The $y$-axis represents the expected value of missing entities over multiple attempts with a trial size of one.
Hence, it corresponds to the likelihood of missing one entity.

\begin{figure}
    \centering
    \includegraphics[width=\textwidth]{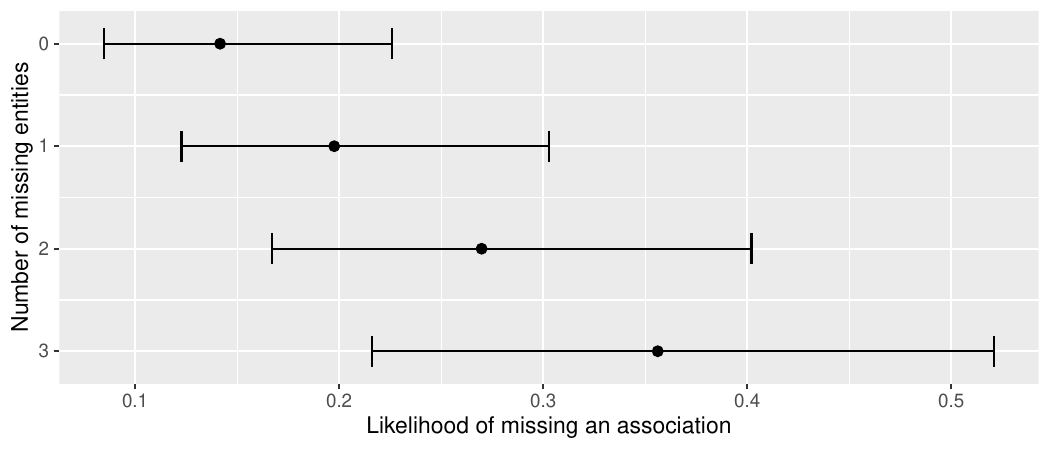}
    \caption{Impact of missing entities on missing associations.}
    \label{fig:missingentities}
\end{figure}

The plot supports the assumption that missing an entity promotes missing an association, which the original experiment did not consider and instead attributed the missing associations fully to the use of passive voice~\cite{femmer2014impact}.
In fact, the strength of the effect of passive voice on missing associations ($\mu_{RQT}^{PV}=0.38$) is similar to the strength of the effect of missing entities on missing associations ($\mu_{E^-}=0.40$).
However, the uncertainty of the impact of passive voice ($\sigma_{RQT}^{PV}=0.32$) is higher than that of missing entities ($\sigma_{E^-}=0.09$).
This means that the effect of missing entities on missing associations is much more reliable than the effect of passive voice on missing associations.

\paragraph{Impact of duration}

\Cref{fig:marginal:duration} visualizes the impact of relative duration (i.e., deviation in duration from the overall average time of creating a domain model in minutes) on the two response variables superfluous entities and wrong associations.
The red estimate shows that the longer a participant took to generate a domain model (relative duration $> 0$), the more likely they were to introduce superfluous entities.
The cyan estimate shows that the shorter time a participant took to generate a domain model (relative duration $< 0$), the more likely they were to connect an association wrongly.

\begin{figure}
    \centering
    \includegraphics[width=\textwidth]{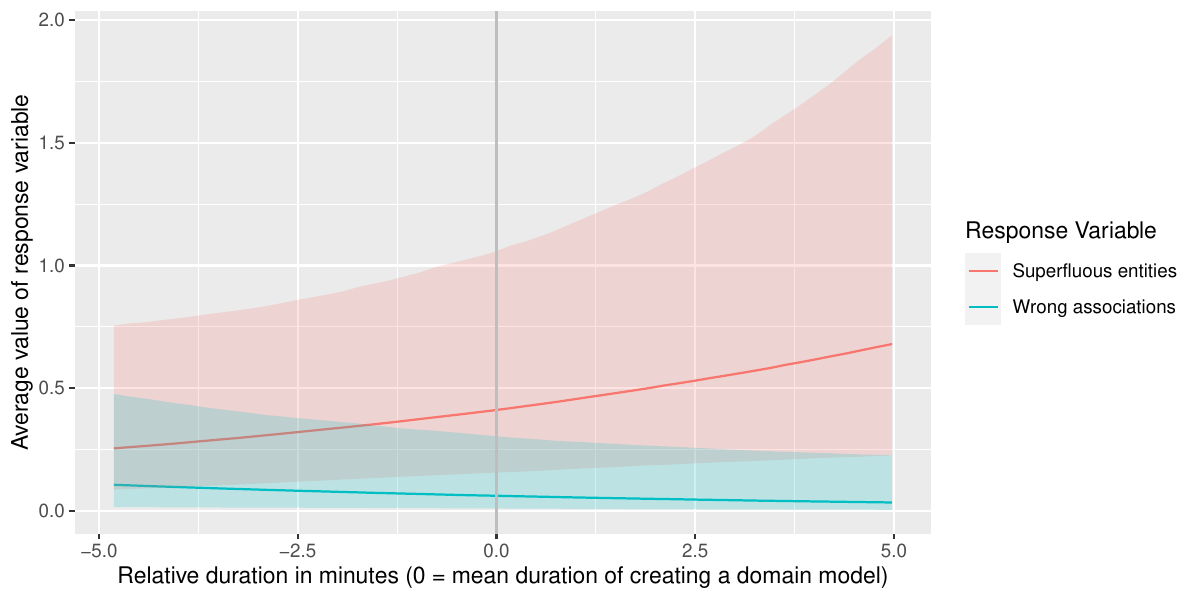}
    \caption{Impact of relative duration on the number of superfluous entities and wrong associations}
    \label{fig:marginal:duration}
\end{figure}

\paragraph{Impact of previous training in modeling}

\Cref{fig:marginal:train} visualizes the impact of prior formal training in modeling on the number of wrong associations in a domain model. 
A participant with prior formal training ($formal = TRUE$) shows a slightly lower likelihood of connecting associations wrongly.
The overlapping confidence intervals do, however, indicate a strong variance of the effect.

\begin{figure}
    \centering
    \includegraphics[width=\textwidth]{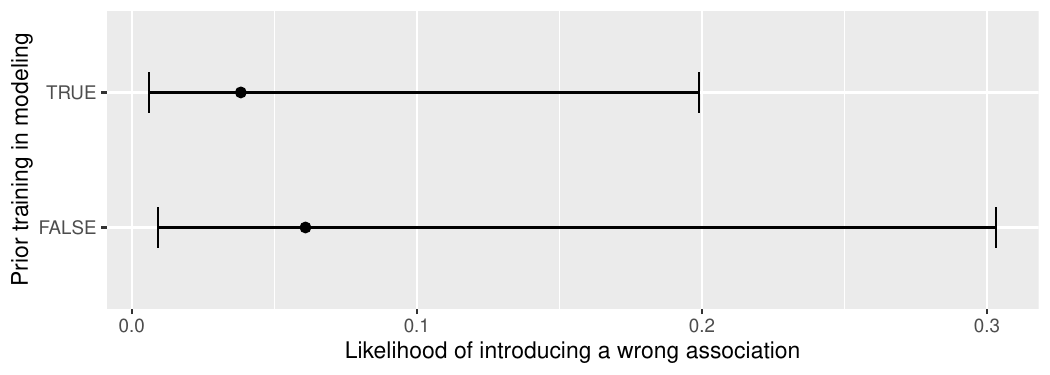}
    \caption{Marginal effect of prior formal training in modeling on the number of wrong associations}
    \label{fig:marginal:train}
\end{figure}

\paragraph{Impact of remaining context factors}

None of the remaining context factors has a stronger effect on any of the response variables than the previously mentioned impact visualized in \Cref{fig:marginal:train}.
This means that the other context factors are neither notable ($\mu>0.4$) nor significant ($\sigma<\mu$).
The replication package contains a detailed summary of all coefficients.

\paragraph{Interaction between domain knowledge and the treatment}

Conditional plots visualize interaction effects between two predictors.
\Cref{fig:interaction:domain} visualizes the interaction effect between domain knowledge about open source and the treatment on the number of wrong associations.
The figure shows that the impact of ambiguous pronouns (cyan whisker plots) on the response variable \textit{number of wrong associations} diminishes the greater the domain knowledge about open source.
For the co-occurrence of ambiguous pronouns and passive voice (purple whisker plots), the effect is less pronounced but symmetrical, i.e., the factor has the strongest impact on the response variable when the domain knowledge is medium.
However, the effect contains high uncertainty when the treatment involves ambiguous pronouns, represented by the large and overlapping confidence intervals (cyan and purple whiskers in \Cref{fig:interaction:domain}).
The collected data does not suffice to support the significance of this effect.

\begin{figure}
    \centering
    \includegraphics[width=\textwidth]{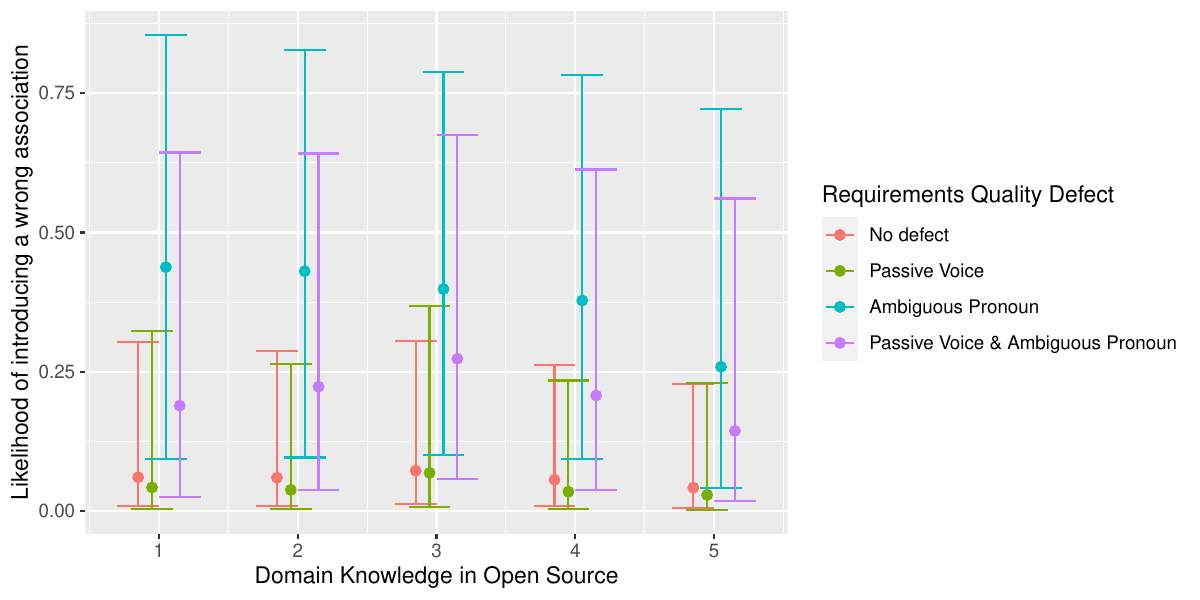}
    \caption{Interaction effect between domain knowledge and the treatment on the number of wrong associations}
    \label{fig:interaction:domain}
\end{figure}

\begin{highlightbox}{Result of Marginal and Conditional Effects}
    Most context factors do not show a significant impact on either the response variables directly or mediate the effect of quality defects. The few context factors that do show an impact are not significant.
\end{highlightbox}

\subsection{Comparison of original with our Study Results}
\label{sec:result:comparison:replication}

{For the part of our study that serves as a conceptual replication, we compare the results of the original study~\mbox{\cite{femmer2014impact}} and its re-analysis~\mbox{\cite{frattini2024second}} with our results~\mbox{\cite{carver2010towards}}.
Regarding $H_0^{PV \rightarrow E^-}$, we obtain conflicting results from the FDA as we reject the null hypothesis while the original study does not, but consistent results from the BDA, as the distribution of the posterior prediction of missing entities is balanced.
We obtain conflicting results for $H_0^{PV \rightarrow A^-}$ from the FDA as we cannot reject it as in the original study.
Our BDA suggests that passive voice has a slight impact on the number of missing associations (25.5\% less and 45.0\% more likely to miss an association).
The result of the BDA does not suppose an effect as strong as the original study, but it does agree with the re-analysis of the original study~\mbox{\cite{frattini2024second}} on a slight impact.
Overall, the results of our BDA agree with the reanalyzed results of the original study.
The variation of study elements (e.g., experimental subjects and objects)~\mbox{\cite{gomez2010replications}} increases the replicability space within the generalizability space~\mbox{\cite{nosek2020replication}} and identifies those elements as non-influential~\mbox{\cite{juristo2011role}} to the original claim.
}

\begin{highlightbox}{Comparison of Studies}
    {The results of the frequentist analyses of the original and our study differ. 
    However, the more thorough Bayesian data analysis agrees with the properly re-analyzed original results.
    Due to the variation of several elements of our study from the original study, the conceptual replication extends the external validity of the original claim that passive voice has only a slight impact on the domain modeling activity.
    }
\end{highlightbox}

\subsection{Comparison of FDA with BDA Results}
\label{sec:result:comparison:bda}

Secondly, we compare the results of our frequentist data analysis with the results of our Bayesian data analysis.
We obtain \textit{consistent results}~\cite{carver2010towards} for $H_0^{RQD \in \{PV, AP, PVAP\} \rightarrow D}$.
Neither the frequentist nor Bayesian analysis suggests an impact of the treatment on the relative duration to create a domain model.

The frequentist analysis rejects $H_0^{RQD \in \{PV, AP, \text{\textit{PVAP}}\} \rightarrow E^-}$, while the Bayesian analysis remains more cautious.
The posterior predictions in \Cref{tab:posteriors} show a tendency towards the treatment having an impact, but with large uncertainty.
Additionally, marginal plots of the Bayesian analysis reveal that the primary role, experience with domain modeling, and education impact the dependent variable.
Both analyses agree that $H_0^{RQD \in \{PV, AP, \text{\textit{PVAP}}\} \rightarrow E^+}$ cannot be rejected, though the Bayesian analysis attributes a tendency of causing superfluous entities to ambiguous pronouns.

The frequentist analysis suggests to reject {$H_0^{RQD \in \{AP, \text{\textit{PVAP}}\} \rightarrow A^-}$}, i.e., ambiguous pronouns and their coexistence with passive voice influence the number of missing associations.
The Bayesian analysis again shows a tendency towards an impact but retains its uncertainty about the effect.
Marginal plots instead emphasize the influence of missing entities on the response variable.

The analyses agree on the impact of ambiguous pronouns on the number of wrong associations and suggest to reject $H_0^{RQD \in \{AP, \text{\textit{PVAP}}\} \rightarrow A^\times}$.
Both the large effect size and the skewed distribution of posterior predictions support the existence of a causal effect of ambiguous pronouns on wrong associations in a domain model.

\begin{highlightbox}{Comparison of Analysis Methods}
    The results of our frequentist analysis differ from our Bayesian analysis: 
    the Bayesian data analysis remains more cautious about several effects suggested by the frequentist analysis.
    The extended casual model attributes part of the effect on the response variable on other independent variables than the treatment.
\end{highlightbox}

\section{Discussion}
\label{sec:discussion}

\Cref{sec:discussion:rqs} answers the research questions.
\Cref{sec:discussion:practice} discusses implications for requirements quality practice and \Cref{sec:discussion:bda} for requirements quality research.
\Cref{sec:discussion:threats} presents the threats to validity. 

\subsection{Answers to research questions}
\label{sec:discussion:rqs}

\subsubsection{Answer to RQ1}
\label{sec:discussion:rqs:1}

\paragraph{RQ1.1: Impact of passive voice.} 
Using passive voice in natural language requirements specifications has a slightly negative effect on the domain modeling activity regarding missing associations.
This finding aligns with the conclusions drawn by the original study by Femmer et al.~\cite{femmer2014impact,frattini2024second}.
However, the Bayesian data analysis emphasizes that both context factors, and especially the number of missing entities, have a significant impact on the number of missing associations as well.
Overall, these results support the claim that passive voice can have a negative impact in specific cases but is overall not a significant factor in subsequent activities depending on the requirement~\cite{femmer2014impact,krisch2015myth}.

\paragraph{RQ1.2: Impact of ambiguous pronouns.} 
The use of ambiguous pronouns has a strong effect on the number of wrong associations in the resulting domain model.
Additionally, using ambiguous pronouns has a slight negative effect on the number of missing and superfluous entities and missing associations.
This confirms the risk of using ambiguous pronouns that have been mainly hypothesized in previous research~\cite{ezzini2022automated} and explains the focus on ambiguity in requirements quality research~\cite{montgomery2022empirical}.
An ambiguous pronoun in a requirements specification has a 44.6\% chance of causing a wrongly connected association in the domain model, limiting the model's correctness and propagating risk to further activities.

\paragraph{RQ1.3: Combined impact.} 
The co-occurrence of passive voice and ambiguous pronouns has a strong effect on the number of wrong associations.
Additionally, it has a slight effect on the number of missing entities and associations.
The impact correlates with but never exceeds the effect of pure, ambiguous pronouns.
This supports the assumption that passive voice does not create any further impact in addition to the effect of ambiguous pronouns.

\subsubsection{Answer to RQ2}
\label{sec:discussion:rqs:2}

\paragraph{RQ2.1: Impact of context factors.} 
Only a small number of context factors included in the study show a notable effect on the response variables.
The duration of the domain modeling activity confirms assumed patterns: 
taking shorter than average increases the chance of missing elements or connecting associations wrongly, taking longer time than average increases the chance of adding superfluous entities.
Prior formal training in modeling shows a slight yet not significant positive effect.

\paragraph{RQ2.2: Mediation of context factors.}
The interaction effect between domain knowledge and the treatment shows that higher domain knowledge can mitigate the negative effect of quality defects on response variables.
In particular: higher domain knowledge reduces the chance of connecting associations incorrectly.
While the effect still exhibits a large variance, this hints at the possibility of compensating quality defects with domain knowledge.

\subsection{Implications for Requirements Quality Practice}
\label{sec:discussion:practice}

The presented results indicate that the negative impact of two requirements quality defects can differ significantly.
When allocating resources toward detecting and removing specific quality defects from requirements specifications, organizations can make informed decisions based on the calculated impact of the respective quality factor.
In our case, we recommend explicitly detecting and resolving ambiguous pronouns, while passive voice is not critical enough to deserve dedicated attention.
This aligns with the common perception in requirements quality research that ambiguity receives the most attention~\cite{montgomery2022empirical} while using passive voice rarely has a tangible impact~\cite{krisch2015myth}.
By filtering requirements writing guidelines for quality factors that have a measurable effect, we expect greater acceptance of requirements quality assessment tools in practice~\cite{franch2020practitioners,phalp2007assessing}.

Additionally, measuring the effect of a quality defect on the relevant attributes of activities that use these requirements allows quantifying it economically~\cite{frattini2023requirements}.
While the cost of a quality defect is hard to determine, a company can quantify the cost of activities' attributes like increased duration.
This economic perspective provides additional decision support for companies when assessing whether it is worth detecting and removing a specific quality defect~\cite{juergens2010much}.

Finally, the potential influence of context factors on the impact of quality defects on affected activities may incentivize organizations to invest in developing these factors.
For example, improving domain knowledge and providing formal modeling training may compensate for quality defects.

\subsection{Implications for Requirements Quality Research}
\label{sec:discussion:bda}

Employing Bayesian data analysis to investigate the impact of requirements quality defects provides sophisticated and sensitive insights necessary to propel requirements quality research~\cite{frattini2023requirements}.
The result of the analysis models both the direction and strength of an impacting factor while retaining information about its certainty.
These insights go beyond the point-wise comparison and binary result of frequentist analyses~\cite{furia2019bayesian}.
The frequentist analysis fails to compare the impact of quality defects on the response variables, as even the calculated effect sizes are similar ($0.79<|ES|<0.93$). 
The Bayesian data analysis, on the other hand, clearly shows that some effects are much stronger (e.g., $H_0^{AP \rightarrow A^\times}$) than others (e.g., $H_0^{PV \rightarrow A^-}$).
{Still, the BDA relies on the causal model expressed in a DAG, statistical assumptions about variable types and their independence, and the validity of constructs.
Therefore, the results obtained via BDA cannot be seen as more valid by design.
However, the BDA is more transparent and allows critical debate---e.g., about the causal assumptions underlying our analysis in \mbox{\Cref{fig:dag}}---which facilitates the incremental improvement of empirical studies.}

Including context factors in the prediction allows comparing the impact of requirements quality with the impact of human and process factors, revealing which causes changes in the response variable.
These context factors can also represent the properties of non-human agents like generative artificial intelligence (GenAI) models which are increasingly employed for RE tasks.
Involving context factors like the version number of a GenAI model, its parameters, its context window, and other factors in empirical studies resembling our approach will allow to investigate which configurations of these models excel at performing their RE task.

Abandoning simple NHSTs for identifying relevant factors of requirements quality and instead opting for a proper framework for causal inference like BDA will increase the likelihood of solving problems with practical relevance~\mbox{\cite{franch2020practitioners}} that justify subsequent tool development~\mbox{\cite{frattini2024nlp4re}}.
Empirical studies with explicit causal assumptions (e.g., visualized as DAGs) and sophisticated analyses will produce context-sensitive evidence that can be synthesized in the common framework of the requirements quality theory~\mbox{\cite{frattini2023requirements}}.
The continuous synthesis of evidence from individual studies in this common framework will produce more reliable and generalizable conclusions~\mbox{\cite{badampudi2019contextualizing,ciolkowski2005accumulation}} and effectively address the lack of empirical insights in requirements quality research~\mbox{\cite{montgomery2022empirical}}.

Using a controlled experiment benefits the investigation of the quality factor~\cite{femmer2014impact}.
The DAG shown in \Cref{fig:dag} visualizes this control, as no other factor influences the treatment in question.
This eliminates spurious associations that could confuse the results~\cite{mcelreath2020statistical}.
On the other hand, the cost of conducting a controlled experiment---especially with participants from industry---cannot be neglected~\cite{sjoberg2003challenges}.
Luckily, statistical causal inference via Bayesian data analysis works equally well with observational data, as shown by Furia et al.~\cite{furia2023towards}.


Finally, Bayesian data analysis allows for incremental improvement of empirical inquiry regarding requirements quality.
The causal assumptions that the DAG makes explicit can be reviewed, discussed, and updated to inform future empirical methods.
Insights derived from Bayesian data analysis can be used as prior knowledge in subsequent analyses, just as we sensibly used previous results~\cite{femmer2014impact} to inform our priors.

Worth noting is that our comparison between FDA and BDA conflates the use of causal frameworks with advanced Bayesian statistics. 
An FDA can also employ causal frameworks that mitigate parts of the shortcomings mentioned in \mbox{\Cref{sec:related:bdainse}}, as previously shown by Furia et al.~\mbox{\cite{furia2023towards}}.
However, frequentist approaches tend to limit their analyses to the treatment and the response variable, disregarding potential context~\mbox{\cite{mund2015does}} or experimental design factors~\mbox{\cite{vegas2015crossover}}.
BDA, on the other hand, entails the use of an explicit causal framework~\mbox{\cite{mcelreath2020statistical,pearl1995bayesian}}, which is why we support the recommendation of abandoning FDA for BDA in SE research~\mbox{\cite{torkar2020missing,furia2019bayesian}}.

\begin{highlightbox}{Implications}
    Quality defects in requirements specifications have a varying impact on affected activities that depend on them.
    Context factors may compensate for this impact but require better metrics to quantify them. 
    Bayesian data analysis provides more fine-grained insights into these effects than frequentist methods.
\end{highlightbox}

\subsection{Threats to Validity}
\label{sec:discussion:threats}

We present and discuss threats that could affect our study based on the guidelines by Wohlin et al.~\cite{wohlin2012experimentation} and extended by the guideline by Vegas et al.~\cite{vegas2015crossover} for the specific threats caused by the use of a crossover design.
The threats to validity are prioritized, considering our work focuses on replicating the first study testing a causal theory predicting the impact of requirements quality factors on downstream development tasks.

\subsubsection{Internal validity}
\label{sec:discussion:threats:internal}

Our design and the blind nature of the experiment avoid the threat to \textit{selection-maturation interaction}. 
Nevertheless, the new settings (i.e., online asynchronous experiment) may have caused a \textit{diffusion or imitation of treatments}---i.e., information may have been exchanged among the participants.
The experiment supervisor monitored the participants to prevent their communication with each other and asked them not to distribute the experimental task and materials.
We acknowledge that \textit{selection} can bias our sample as volunteers are generally more motivated to perform in an experimental task~\cite{baltes2022sampling}.

The crossover design emits additional threats to validity~\cite{vegas2015crossover}.
We mitigate the \textit{learning by practice} effect---i.e., participants getting better when repeating the experimental task---in three ways:
Firstly, we disperse the learning effect evenly at design time by randomizing the sequences of treatments.
Secondly, we include a warm-up object to get participants used to the task and tool but exclude that data from the analysis.
Thirdly, we include the \textit{period} variable as a predictor to factor out the learning effect during the analysis.
We avoid the threat of \textit{copying} by prohibiting communication among participants and using experimental objects where solutions cannot be copied from one task to another.

The threat of \textit{optimal sequence} describes the risk that there is a sequence in which the treatment is applied, which optimizes the participants' performance in deriving domain models.
We cannot block this threat at analysis time as the sequence and participant IDs are highly correlated.
This is because we could---in all but one case---assign only one participant ($n_p=25$) to each sequence ($n_s=n_r!=4!=24$).
Because of this strong correlation, the Bayesian model is incapable of distinguishing between the impact of the sequence ($\beta_{seq}$) from the within-participant variance ($\alpha_{PID}$)~\cite{vegas2015crossover}.
More participants per sequence would have been necessary to block the threat of an optimal sequence, but these were unavailable to us.

Finally, we address the threat of \textit{carryover}---i.e., the change of the impact caused by the period in which the treatment was applied---at analysis time by including the term $period*treatment$ in the predictors.
This way, the carryover effect is factored out from the impact of the treatment and analyzable from the posterior distributions.

\subsubsection{Conclusion validity}
\label{sec:discussion:threats:conclusion}

{We addressed the \textit{reliability of measures} threat by creating and disclosing evaluation guidelines and peer-reviewing the extraction of the dependent variables from the collected domain models.
Despite the acceptable inter-rater agreement score, an in-depth qualitative evaluation of the remaining disagreements may be beneficial to further improve the evaluation instrument and, therefore, the reliability of the results.}
We addressed the \textit{random heterogeneity of the subjects} by a design in which each participant acts as their own control group. 

Moreover, we focused on including and analyzing context variables related to the participants' experience.
Our sample of participants is not representative of all context factors. 
Consequently, our Bayesian data analysis cannot identify all causal effects of some context factors.
However, by including them in the causal considerations, the effect of the factors is isolated from the potential confounding variables~\cite{leven2022broken}.

The conclusion validity of our study is strengthened by applying two different data analysis approaches and comparing their results.
The data analysis suffers from the threat of \textit{low statistical power} when it comes to evaluating interaction effects, as reliably identifying them requires a larger sample size~\cite{gelman2018interaction}.
We limit the number of interaction effects considered in our models and discuss the uncertainty around the coefficient estimates to minimize this threat.

{The analysis can suffer from \textit{violated assumptions of statistical tests}.
Modeling the number of missing entities and associations as binomial distributions implies the independence of each event, i.e., that each missing entity and association is independent of all other missing entities and associations.
While we did not observe any cascading, i.e., dependent, defects, their independence remains only assumed.
}

\subsubsection{Construct validity}
\label{sec:discussion:threats:construct}

Our study can suffer from \textit{mono-operation bias} as we focus only on a subset of quality factors that can potentially exist~\cite{montgomery2022empirical,frattini2022live}. 
Nevertheless, our goal with this replication is to extend the initial quality factor of passive voice reported by Femmer et al.~\cite{femmer2014impact} to a second one---ambiguous pronouns---which is widespread as indicated by the literature~\cite{frattini2022live,montgomery2022empirical}.

Similarly, a \textit{confounding of constructs and level of constructs} could influence the outcomes of our study, for example, the presence of several ambiguous pronouns or passive voice sentences rather than their binary presence or absence from a specification. 
Further replications, focusing on improving construct validity, should include several levels of each treatment.

\textit{Mono-method bias} is a potential threat to construct validity---i.e., we measured the dependent variables using a single type of measurement, inspired by the original study. 
However, the measurements were based on a pre-defined protocol and peer-reviewed. 
Our study may result in a \textit{restricted generalizability across constructs} since the presence or absence of the different quality factors could result in side effects for other interesting outcomes we did not measure (e.g., comprehensibility or maintainability of the specification). 

Among the social threats to construct validity, we acknowledge that \textit{hypothesis guessing} may have taken place since the participants could try to guess the concrete goal of the experimental task based on the invitation text and material provided during the sessions. 
Nevertheless, we used the same text and phrasing to invite all participants and the same material during the experimental task.
\textit{Evaluation apprehension} could have played a role since some participants are students at the authors' institution. 
However, students did not receive rewards (e.g., extra grade points) for participating in the experiment, and they received their course grades before the start of the experiment. 

Moreover, our study can suffer from an \textit{inadequate preoperational explication of constructs} as we did not validate our context factors.
For example, we are unable to provide any proof that the self-reported number of years spent in RE adequately represents the latent variable of experience in RE beyond educated guesses and relying on comparable practices in our scientific community~\cite{bogner2023restful}.
To improve the construct validity, separate studies investigating the adequacy of these measurements in representing their constructs are necessary.
{This particularly impacts our decision to replace the binary distinction of participants by type (students versus practitioners) with more fine-grained variables like experience and domain knowledge.
While our study supports the feasibility of this step on an analytical level, we cannot prove its validity on a conceptual level.
We encourage investigating the feasibility of variables to represent individual skills to improve the construct validity of studies considering this impact~\mbox{\cite{salman2015students}}.
}

Finally, a variable of the selected population that may interact with the treatment that we did not analyze is the \textit{language skill} of participants.
Arguably, skills in the English language influence the ability to comprehend and process the experimental objects and, therefore, may impact the response variables.
We were unable to measure this variable properly given that all participants scored the same on the Common European Framework of Reference for Languages
(CEFR)~\cite{martyniuk2006common} (i.e., non-native, fluent English speakers).
While the threat is minimized in our study due to the comparable language level of participants, future studies should develop measurement instruments for this construct and involve this variable in such causal queries.

\subsubsection{External validity}
\label{sec:discussion:threats:external}

The main threat to the external validity of this study is the \textit{interaction of setting and treatment} as the size and the complexity of the selected specifications, despite being sampled from a real-world data set, might not be representative of the industrial practice. 
{Using Google Docs as the modeling tool is not fully representative of real-world practices.
Given that it was appropriate and sufficient for the experimental task, however, renders this as an opportunity for improving the realism of the experiment in future studies rather than a threat to validity.}

{There may be the threat of \textit{interaction of selection and treatment}, as some participants reported no modeling experience or training.
These deficiencies might influence the results and render a subset of the participants as non-representative of our target population.
We attempted to mitigate this threat via comprehensive instructions and including a warm-up phase in the experiment.}

\section{Conclusion}
\label{sec:conclusion}

Requirements quality research lacks empirical evidence and research strategies to advance beyond proposing and following normative rules with unclear impact~\cite{frattini2022live} to better understanding and solving problems relevant to practice~\cite{franch2017practitioners,franch2020practitioners}.
In the scope of our study, we {conducted a} controlled experiment on the impact of requirements quality defects on subsequent activities.
We demonstrated a method of evaluating data collected through a controlled experiment using a crossover design with Bayesian data analysis. 
We showed the impact (1) of requirements quality defects varies and (2) may be mediated by context and confounding factors.
{The part of our study that serves as a conceptual replication strengthens the claims of the re-analyzed original study~\mbox{\cite{femmer2014impact,frattini2024second}} that passive voice only has a slight impact on missing associations from domain models.}

We can confidently support the recommendation of SE researchers to adopt Bayesian data analysis to improve causal reasoning and inference~\cite{furia2019bayesian,furia2022applying,torkar2020missing}, which will propel requirements quality research.
This shift requires focusing on problems such as scrutinizing the explicit causal assumptions of a DAG, visualizing requirements quality impact, evolving prior knowledge about their impact, and comparing models concerning their predictive power.

We envision that adopting sophisticated statistical tools like Bayesian data analysis and the focus of empirical studies on investigating the impact of requirements quality defects will steer requirements quality research in a relevant and effective direction.
{Explicit causal assumptions and sophisticated data analyses will produce empirical evidence which can be more easily synthesized to more reliable and generalizable conclusions~\mbox{\cite{ciolkowski2005accumulation}} in a common framework~\mbox{\cite{frattini2023requirements}}.}
We hope that the documentation of this study inspires fellow researchers to adopt our method and tools for replication.

\begin{acknowledgements}
    This work was supported by the KKS foundation through the S.E.R.T\@. Research Profile project at Blekinge Institute of Technology.
    We are deeply grateful to Parisa Yousefi from Ericsson AB for recruiting practitioners to the experiment.
    We further thank the reviewers for their tremendous effort that significantly improved this manuscript.
\end{acknowledgements}

\section*{Conflict of interest}

The authors declare that they have no conflict of interest.

\section*{Data Availability Statement}

All supplementary material, including protocols and guidelines for data collection and extraction, the raw data, analysis scripts, figures, and results, are available in our replication package~\cite{frattini2024replication}.
The replication package is available on GitHub at \url{https://github.com/JulianFrattini/rqi-proto} and archived on Zenodo at \url{https://doi.org/10.5281/zenodo.10423666}.

\bibliographystyle{spmpsci}
\bibliography{references}

\end{document}